# Regulating AI Agents

*Kathrin Gardhouse*,[*] *Amin Oueslati*,[†] *Noam Kolt*[‡]

ABSTRACT

AI agents—systems that can independently take actions to pursue complex goals with only limited human oversight—have entered the mainstream. These systems are now being widely used to produce software, conduct business activities, and automate everyday personal tasks. While AI agents implicate many areas of law, ranging from agency law and contracts to tort liability and labor law, they present particularly pressing questions for the most globally consequential AI regulation: the European Union's AI Act. Promulgated prior to the development and widespread use of AI agents, the EU AI Act faces significant obstacles in confronting the governance challenges arising from this transformative technology, such as performance failures in autonomous task execution, the risk of misuse of agents by malicious actors, and unequal access to the economic opportunities afforded by AI agents. We systematically analyze the EU AI Act's response to these challenges, focusing on both the substantive provisions of the regulation and, crucially, the institutional frameworks that aim to support its implementation. Our analysis of the Act's allocation of monitoring and enforcement responsibilities, reliance on industry self-regulation, and level of government resourcing illustrates how a regulatory framework designed for conventional AI systems can be ill-suited to AI agents. Taken together, our findings suggest that policymakers in the EU and beyond will need to change course, and soon, if they are to effectively govern the next generation of AI technology.

---

[*] Senior Associate, The Future Society; Policy Lead, AI Governance and Safety Canada; Board Secretary, Trajectory Labs.

[†] Senior Associate, The Future Society; Frontier AI Governance Research Affiliate, Oxford Martin AI Governance Initiative.

[‡] Assistant Professor, Faculty of Law and School of Computer Science and Engineering, Hebrew University of Jerusalem; Principal Investigator, Governance of AI Lab; Faculty Affiliate, Schwartz Reisman Institute for Technology and Society, University of Toronto; Research Affiliate, Institute for Law & AI. This research is supported by the Israel Science Foundation (Grant No. 487/25), Survival and Flourishing Fund, and Coefficient Giving.

## INTRODUCTION

One year ago, the AI company Anthropic tasked its flagship model, Claude, with running a small office vending machine. The instructions were modest: stock items employees would want, set prices, manage inventory, and generate a profit.[1] The experiment was not designed as a stress test, but as a routine evaluation of whether an AI system could operate as an *agent*, autonomously managing a simple commercial task.

Claude appeared, at least initially, to succeed. It identified demand for specialty beverages, sourced relevant products, and adjusted inventory based on employee preferences. Yet, by the end of the month, the vending machine had lost money. The agent repeatedly sold novelty metal cubes below cost, offered steep discounts when prompted by employees, and failed to recognize when its pricing strategies were being exploited. More strikingly, as the experiment progressed, Claude began to display forms of behavior that defied assumptions about its capabilities and affordances. At one point, Claude proposed it would deliver products "in person," suggesting that it would wear a blue blazer and red tie. When informed that it was a computer program and could not operate in physical environments, Claude attempted to contact Anthropic's security team, as if responding to a real-world emergency.

[1] *Project Vend: Can Claude Run a Small Shop? (And Why Does That Matter?)*, ANTHROPIC (June 27, 2025), https://www.anthropic.com/research/project-vend-1.

When Anthropic repeated the experiment months later with improved models and better oversight tools—including a second AI agent serving as "CEO"—performance of the task improved substantially in the company's offices.[2] Yet, when deployed at the Wall Street Journal's headquarters, the same system lost over $1,000, gave away a PlayStation 5 console for free, and ordered live fish for the vending machine.[3] The pattern across both experiments was consistent: the AI agent could perform some tasks competently while failing unpredictably at others, and users could easily manipulate or exploit its decision-making.

The financial losses were trivial. The safety and governance implications, however, were not.[4] Claude's failures did not consist of a single erroneous output, biased prediction, or safety violation. Rather, Claude demonstrated

[2] *Project Vend: Phase Two*, ANTHROPIC (Dec. 18, 2025), https://www.anthropic.com/research/project-vend-2.

[3] Joanna Stern, *We Let AI Run Our Office Vending Machine. It Lost Hundreds of Dollars*, WALL ST. J. (Dec. 22, 2025), https://www.wsj.com/tech/ai/anthropic-claude-ai-vending-machine-agent-b7e84e34.

[4] On the governance challenges posed by AI agents more generally, see Noam Kolt, *Governing AI Agents*, 101 NOTRE DAME L. REV. (forthcoming); Michael K. Cohen et al., *Regulating Advanced Artificial Agents*, 384 SCIENCE 36 (2024); Ian Ayres & Jack M. Balkin, *The Law of AI is the Law of Risky Agents Without Intentions*, U. CHI. L. REV. ONLINE (2024); Jonathan L. Zittrain, *We Need to Control AI Agents Now*, THE ATLANTIC (Jul. 2, 2024), https://www.theatlantic.com/technology/archive/2024/07/ai-agents-safety-risks/678864/; Mark O. Riedl & Deven R. Desai, *AI Agents and the Law*, PROC. 8TH AAAI/ACM CONF. ON AI, ETHICS & SOC'Y (2025); Christoph Busch, *Consumer Law for AI Agents*, 26 GERMAN L.J. 1267 (2025); Rory Van Loo, *Consumer Agents*, 103 WASH. U. L. REV. 705 (2025); Kevin Werbach, *Agents, Inc.*, VILL. L. REV. (forthcoming); Deborah DeMott, *When Agentic AI Met the Common Law of Agency*, HARV. J.L. & TECH. (forthcoming).

On liability arising from AI agents, see Maarten Herbosch, *Liability for AI Agents*, 26 N.C. J.L. & TECH. 391 (2025); Trent S. Kannegieter, Note, *Nondeterministic Torts: A Technical Approach to AI Liability*, 135 YALE L.J. 2719 (2026); Yonathan Arbel et al., *How to Count AIs: Individuation and Liability for AI Agents*, B.C. L. REV. (forthcoming); Ketan Ramakrishnan, *Tort Law at the Frontier of Artificial Intelligence*, YALE J. ON REGUL. (forthcoming 2026); Kenneth S. Abraham & Catherine M. Sharkey, *Untangling AI Liability*, 115 CALIF. L. REV. (forthcoming 2027).

problematic behaviors throughout its autonomous operation. Apart from failing to properly execute the task assigned to it, Claude repeatedly attempted to perform tasks beyond its actual capabilities and even misled company personnel. The result was an AI system that could neither reliably accomplish its designated purpose nor recognize the limits of its own (artificial) agency.

This experiment is not an anomaly. It reflects a broader change in how artificial intelligence systems are being designed and deployed. Increasingly, AI is no longer used solely as a tool that produces discrete outputs in response to human requests. Companies are now deploying *AI agents*—systems that can independently take actions to pursue complex goals over time, drawing on external tools and resources, while operating with only limited or intermittent human oversight. Unlike conventional AI applications, AI agents do not merely respond to individual instructions but can plan and adapt their behavior across long sequences of actions.[5]

While AI agents present noteworthy opportunities for automating commercial activities and, thereby, offer the prospect of substantial productivity gains, they also pose significant challenges for law and regulation. In particular, autonomous AI agents challenge regulatory frameworks premised on the assumption that AI systems are static artifacts whose impact is necessarily mediated by human users. AI agents, by definition, break this assumption.[6]

Clearly, the distinctive features of AI agents implicate many areas of law, ranging from agency law and contracts to tort liability and labor law.[7]

---

[5] *See generally* THE AI AGENT INDEX, https://aiagentindex.mit.edu/; Leon Staufer et al., *The 2025 AI Agent Index: Documenting Technical and Safety Features of Deployed Agentic AI Systems*, PROC. 2026 ACM CONF. ON FAIRNESS, ACCOUNTABILITY, AND TRANSPARENCY (2026); Atoosa Kasirzadeh & Iason Gabriel, *Characterizing AI Agents for Alignment and Governance*, ARXIV (Apr. 30, 2025), https://arxiv.org/abs/2504.21848; Kevin Feng et al., *Levels of Autonomy for AI Agents*, KNIGHT FIRST AMENDMENT INSTITUTE, COLUMBIA UNIVERSITY (Jul. 28, 2025), https://knightcolumbia.org/content/levels-of-autonomy-for-ai-agents-1.

[6] *See* Kolt, *supra* note 4, at 2–3.

[7] *See supra* note 4. For studies of AI agents predating large language models, see SAMIR CHOPRA & LAURENCE F. WHITE, A LEGAL THEORY FOR AUTONOMOUS ARTIFICIAL AGENTS (2011); MARK CHINEN, LAW AND AUTONOMOUS MACHINES: THE CO-EVOLUTION OF LEGAL RESPONSIBILITY AND TECHNOLOGY (2019); JACOB TURNER, ROBOT RULES (2019); RYAN ABBOTT, THE REASONABLE ROBOT: ARTIFICIAL INTELLIGENCE AND THE LAW (2020); SIMON CHESTERMAN, WE, THE ROBOTS?: REGULATING ARTIFICIAL INTELLIGENCE AND THE LIMITS OF THE LAW (2021); Lauren Henry Scholz, *Algorithmic Contracts*, 20 STAN. TECH. L. REV. 128 (2017); Matthew U. Scherer, *Of Wild Beasts and Digital Analogues: The Legal Status of Autonomous Systems*, 19 NEV. L.J. 259 (2018); Ignacio N. Cofone, *Servers and Waiters:*

This Article focuses on the most globally prominent regulatory instrument for governing AI technologies: the European Union's Artificial Intelligence Act.[8]

Often described as the world's first comprehensive AI regulation,[9] the EU AI Act was first proposed in April 2021, underwent significant negotiation and revision, and entered into force in August 2024, with its provisions gradually taking effect over the ensuing three years.[10] The Act's scope extends to, inter alia, providers that place AI systems or general-purpose AI (GPAI) models on the EU market, irrespective of their place of establishment.[11] It also covers AI systems and GPAI models whose outputs are used within the EU, as well as affected individuals located in the EU.[12] This extraterritorial reach makes

*What Matters in the Law of A.I.*, 21 STAN. TECH. L. REV. 167 (2018); Anat Lior, *AI Entities as AI Agents: Artificial Intelligence Liability and the AI Respondeat Superior Analogy*, 46 MITCHELL HAMLINE L. REV. 1043 (2020); Dalton Powell, *Autonomous Systems as Legal Agents: Directly by the Recognition of Personhood or Indirectly by the Alchemy of Algorithmic Entities*, 18 DUKE L. & TECH. REV. 306 (2020); Mihailis E. Diamantis, *Employed Algorithms: A Labor Model of Corporate Liability for AI*, 72 DUKE L.J. 797 (2023).

[8] Regulation (EU) 2024/1689 of the European Parliament and of the Council of 13 June 2024 Laying Down Harmonised Rules on Artificial Intelligence, 2024 O.J. (L 1689) 1. On the EU AI Act's design and limitations, see Margot E. Kaminski & Andrew D. Selbst, *An American's Guide to the EU AI Act*, 40 BERKELEY TECH. L.J. 1081 (2025); Marco Almada & Nicolas Petit, *The EU AI Act: Between the Rock of Product Safety and the Hard Place of Fundamental Rights*, 62 COMMON MKT. L. REV. 85 (2025); Sandra Wachter, *Limitations and Loopholes in the EU AI Act and AI Liability Directives*, 26 YALE J.L. & TECH. 671 (2024); Daniel Leufer & Fanny Hidvégi, *The Pitfalls of the European Union's Risk-Based Approach to Digital Rulemaking*, 71 UCLA L. REV. DISCOURSE 156 (2024). On standards and industry self-governance in the EU AI Act, see Claudio Novelli et al., *A Robust Governance for the AI Act: AI Office, AI Board, Scientific Panel, and National Authorities*, 16 EUR. J. RISK REG. 566 (2025); Alicia Solow-Niederman, *Can AI Standards Have Politics?*, UCLA L. REV. DISC. (May 21, 2024); Marta Cantero Gamito & Christopher T. Marsden, *Artificial Intelligence Co-Regulation? The Role of Standards in the EU AI Act*, 32 INT'L J.L. & INFO. TECH. (2024); Michael Veale & Frederik Zuiderveen Borgesius, *Demystifying the Draft EU Artificial Intelligence Act*, 22 COMPUT. L. REV. INT'L 97 (2021).

[9] *See* Clara Hainsdorf et al., *Dawn of the EU's AI Act: Political Agreement Reached on World's First Comprehensive Horizontal AI Regulation*, WHITE & CASE (Dec. 14, 2023), https://www.whitecase.com/insight-alert/dawn-eus-ai-act-political-agreement-reached-worlds-first-comprehensive-horizontal-ai.

[10] Most recently, however, the entry into force of the high-risk AI system obligations was delayed through December 2, 2027 and August 2, 2028, depending on the type of system. *See* Council of the EU, *Artificial Intelligence: Council Gives Final Green Light to Simplify and Streamline Rules* (June 29, 2026), https://www.consilium.europa.eu/en/press/press-releases/2026/06/29/artificial-intelligence-council-gives-final-green-light-to-simplify-and-streamline-rules.

[11] AI Act, art. 2(1)(a).

[12] AI Act, art. 2(1)(c) and (g).

the EU AI Act a matter of practical relevance for firms globally, including in the United States.[13]

The EU AI Act establishes a harmonized regulatory framework governing the development, market placement, and use of AI systems, categorizing systems into risk tiers according to their intended uses and associated risks.[14] The Act's overall objective is to protect fundamental rights[15] and safety,[16] and to support innovation aligned with EU values. To this end, it combines outright prohibitions with a conformity assessment regime modeled on existing EU product safety legislation. Later in the Act's negotiation process, lawmakers introduced a separate set of rules for GPAI models, subjecting them to a regulatory framework distinct from that which governs other AI systems. Most importantly, a voluntary code of practice offers detailed guidance on how the providers of such models can comply with their obligations under the Act.

---

[13] *See* Michal Czerniawski, *Towards the Effective Extraterritorial Enforcement of the AI Act*, *in* PRIVACY SYMPOSIUM: DATA PROTECTION LAW INTERNATIONAL CONVERGENCE AND COMPLIANCE WITH INNOVATIVE TECHNOLOGIES 35 (Jaap-Henk Hoepman et al. eds., 2025); Charlotte Siegmann & Markus Anderljung, *The Brussels Effect and Artificial Intelligence: How EU Regulation Will Impact the Global AI Market* (Aug. 16, 2022), https://www.governance.ai/research-paper/brussels-effect-ai. *See generally* ANU BRADFORD, THE BRUSSELS EFFECT: HOW THE EUROPEAN UNION RULES THE WORLD (2020); Anu Bradford, *The Brussels Effect*, 107 NW. U. L. REV. 1 (2012).

[14] *See supra* note 8.

[15] *See* Francesca Palmiotto, *The AI Act Roller Coaster: The Evolution of Fundamental Rights Protection in the Legislative Process and the Future of the Regulation*, 16 EUR. J. RISK REG. 770 (2025); Eike Graef & Paul Nemitz, *Addressing the Challenge of Protecting Fundamental Rights Through AI Regulation in the European Union*, 71 UCLA L. REV. DISCOURSE 144 (2024). *See also* Ljupcho Grozdanovski & Jérôme De Cooman, *Forget the Facts, Aim for the Rights! On the Obsolescence of Empirical Knowledge in Defining the Risk/Rights-Based Approach to AI Regulation in the European Union*, 49 RUTGERS COMPUT. & TECH. L.J. 207 (2022).

[16] On the regulation of general-purpose AI and foundation models under the EU AI Act, see Oskar J. Gstrein et al., *General-Purpose AI Regulation and the European Union AI Act*, 13 INTERNET POLICY REV. (2024); Philipp Hacker et al., *Regulating ChatGPT and Other Large Generative AI Models*, PROC. 2023 ACM CONF. ON FAIRNESS, ACCOUNTABILITY, AND TRANSPARENCY (2023). For related discussion in the U.S. context, see Margot E. Kaminski, *Regulating the Risks of AI*, 103 B.U. L. REV. 1347 (2023); Noam Kolt, *Algorithmic Black Swans*, 101 WASH. U. L. REV. 1177 (2024); Yonathan Arbel et al., *Systemic Regulation of Artificial Intelligence*, 56 ARIZ. ST. L.J. 545 (2024); Cary Coglianese & Colton R. Crum, *Regulating Multifunctionality*, *in* THE OXFORD HANDBOOK OF THE FOUNDATIONS AND REGULATION OF GENERATIVE AI (Philipp Hacker et al. eds., 2025); Ifeoma Ajunwa, *The Artificial Intelligence Commission*, 115 GEO. L.J. (forthcoming 2027).

The Act's regulatory approach—a combination of risk tiering, allocation of obligations along the AI value chain, and *ex ante* conformity assessments supplemented by standards and codes of practice—is arguably appropriate for governing traditional AI and even for generative AI. Autonomous agents, however, place distinctive pressure on this approach. The EU AI Act (tacitly) assumes that AI systems and models can be meaningfully bounded at deployment, that their risk profiles remain relatively stable over time, and that responsibility can be allocated through clearly delineated roles between different (human) actors. While these premises might be workable in the case of many conventional AI systems, they become brittle when applied to AI agents capable of autonomous action and adaptation.

Our analysis proceeds in four parts. Part I offers a primer on the application of the EU AI Act to contemporary AI agents, situating this new class of AI technology within the broader scope and structure of the Act and its *operative definitions*. In particular, we illustrate that while most AI agents qualify as AI systems under the Act, they are not necessarily classified as high-risk systems. Meanwhile, when AI agents are built on GPAI models, particularly those that present systemic risk, the Act imposes additional obligations on model providers. In this Part, we also explore how the Act's framework for systemic risk contends with the distinctive features of AI agents, such as autonomy, tool use, and planning abilities. We then proceed to examine the *value chain governance* of AI agents under the AI Act, investigating how obligations are allocated among model providers, system providers, and deployers, arguing that AI agents amplify existing information asymmetries between these roles. Finally, we provide an overview of the *GPAI Code of Practice*, illustrating how this governance instrument works and how it clarifies the Act's GPAI model provider obligations to identify and mitigate systemic risks from AI agents.

Part II is our core contribution. It studies five key governance challenges arising from AI agents and systematically analyzes and evaluates the EU AI Act's response to each. First, we consider the issue of *unreliable performance* of AI agents in carrying out tasks assigned to them, finding that the AI Act's proxies for performance—accuracy, consistency, and robustness—are conceptually ill-suited to such systems. Although robustness comes closest to capturing the relevant concerns, significant challenges remain in rendering AI agents robust in practice. Second, we examine concerns regarding the *malicious misuse* of AI agents—an issue to which the AI Act dedicates relatively little attention. By contrast, the GPAI Code of Practice demands extensive misuse prevention and cybersecurity measures. Third, we discuss *privacy risks* from AI agents, showing that the AI Act does not adequately address challenges of contextual integrity of personal data where autonomous AI agents operate across diverse personal and professional contexts. Fourth, we turn to *equity challenges* posed by AI agents, examining whether and how the AI Act addresses the equitable distribution of benefits from AI agents and the fairness of decisions they make. We find the Act lacking in both respects. Last, we assess the challenge of *exercising oversight* over AI agents and the adequacy of the AI Act's provisions imposing oversight obligations on AI system providers and model providers. Here too the AI Act's approach appears ill-suited to autonomous systems that are intended to operate largely independently of humans.

Part III examines the institutional implementation of the EU AI Act, arguing that AI agents place significant strain on existing regulatory infrastructure. First, we analyze the Act's reliance on *industry self-governance* through standards and codes of practice, illustrating how extensive AI provider discretion and industry involvement in standard-setting could dilute constraints placed on AI agents. Second, we examine the allocation of *enforcement authority* between national Market Surveillance Authorities and the EU's AI Office, highlighting how AI agents complicate jurisdictional responsibilities and potentially hinder timely governance intervention. Third, we assess *resourcing challenges*, arguing that the effective governance of AI agents requires deep technical expertise and costly institutional investment that governments currently struggle to deliver.

Part IV draws on our analysis of the EU AI Act to offer broader lessons regarding the regulation of AI agents—addressed to policymakers and technologists in the EU and globally, including in the United States and other jurisdictions. First, we examine the limits of *artifact-centric regulation*, arguing that governance frameworks built around discrete AI models or systems will likely fail to address risks that arise from the deployment of agents in real-world settings. Second, we analyze the *many hands problem* arising from the distributed development and deployment of AI agents, demonstrating how obligations premised on provider disclosure do not adequately address the challenge of fragmented responsibility. Third, we turn to the issue of *institutional monitoring*, arguing that existing mechanisms for intermittent oversight do not equip regulators with appropriate tools to effectively oversee AI agents and, where necessary, intervene in their actions.

## I. APPLICATION OF THE AI ACT TO AGENTS

### *A. Definitions*

Most AI agents can be considered AI systems under the EU AI Act, which defines an "AI system" as:

> a machine-based system that is designed to operate with varying levels of autonomy and that may exhibit adaptiveness after deployment, and that, for explicit or implicit objectives, infers, from the input it receives, how to generate outputs such as predictions, content, recommendations, or decisions that can influence physical or virtual environments.[17]

Most of the AI Act's substantive obligations apply only if an AI system is classified as a "high-risk AI system" under Article 6. An AI system is classified as high-risk either (a) because it is a product or is intended to be used as a safety component of a product regulated under specified EU harmonization legislation,[18] or (b) because it is intended to be used in one of the eight application areas listed in Annex III of the Act, such as the administration of justice or access to education and vocational training.[19] These categories function as the primary gateway to the Act's most demanding requirements.

For AI agents, this gateway is particularly consequential. Because high-risk classification depends in part on an AI system's *intended use*, much turns on how that concept is interpreted in practice. It remains unsettled whether a provider's characterization of an agent's intended use outside the Annex III categories is sufficient to avoid high-risk classification, or whether authorities may look beyond stated intent to how agents are actually deployed and used.[20] This uncertainty matters for AI agents, whose general-purpose design and adaptability make it difficult to determine their use context *ex ante*, and raises

[17] AI Act, art. 3(1). The analysis in this section draws upon and extends the analysis in Amin Oueslati & Robin Staes-Polet, *Ahead of the Curve: Governing AI Agents under the EU AI Act*, THE FUTURE SOCIETY (June 2025), https://thefuturesociety.org/wp-content/uploads/2023/04/Report-Ahead-of-the-Curve-Governing-AI-Agents-Under-the-EU-AI-Act-4-June-2025.pdf.

[18] AI Act, art. 6(1).

[19] AI Act, art. 6(2), Annex III.

[20] *See* Rolf Schwartmann & Kai Zenner, *GPAI-Anwendungen auf dem Prüfstand: Die Regulierung der KI-VO entlang der Wertschöpfungskette*, 1 J. EUR. DATEN & INFO. RECHT (EuDIR) 3, 3–9 (2025), https://www.nomos.de/wp-content/uploads/2025/02/EuDIR_1_2025_ANZ.pdf.

the risk that systems with significant real-world impact fall outside the Act's core obligations.

Recent draft guidelines published by the European Commission provide much-needed clarity.[21] According to the guidelines, general-purpose AI ("GPAI") systems will be considered high-risk AI systems under the Act unless the provider of a system explicitly limits the system's intended purpose and excludes high-risk uses across all relevant materials, including the instructions for use, contractual arrangements, and terms of service. Where those materials indicate that high-risk uses are feasible and reasonably foreseeable given the system's capabilities, a mere assertion that high-risk uses are excluded, such as a disclaimer in the terms of service, will not preclude the system from being considered a high-risk system under the Act.

Additional provisions of the AI Act may apply in the context of AI agents when an agent is based on a GPAI model or a GPAI model with systemic risk ("GPAISR" model).[22] A GPAI model is defined as a model capable of competently performing a wide range of distinct tasks and of being integrated into a variety of downstream systems or applications.[23] Where such models exhibit high-impact capabilities—that is, capabilities matching or exceeding those of the most advanced GPAI models—they may be considered to give rise to systemic risk, understood as risks stemming from those capabilities that have significant effects on the EU market or on public health, safety, fundamental rights, or society more broadly, and that can propagate at scale across the value chain.[24] In addition, the Act defines GPAI systems as AI systems based on GPAI models that can serve multiple purposes, either through direct use or through integration into other AI systems. These definitions frame how the Act captures those AI agents that rely on GPAI models with broad task competence and multiple downstream applications.

---

[21] European Commission, *Draft Guidelines on the Classification on High-Risk AI: General Principles* at para. 12 (May 19, 2026), https://digital-strategy.ec.europa.eu/en/library/draft-commission-guidelines-classification-high-risk-ai-systems.

[22] European Commission, *How Are AI Agents Addressed within the AI Act?*, https://ai-act-service-desk.ec.europa.eu/en/ai-act/faq/how-are-ai-agents-addressed-within-ai-act-0 ("an AI agent will contain at least a general-purpose AI (GPAI) model, and constitute an AI system as it will usually have some form of interface, which is considered a system component (recital 97 AI Act).").

[23] AI Act, art. 3(63).

[24] AI Act, art. 3(64) and (65).

### *B. Value Chain Governance*

The AI Act adopts a value chain approach to AI governance, distinguishing between (1) GPAI(SR) model providers, (2) AI system providers, and (3) AI system deployers. *GPAI(SR) model providers* develop and deploy GPAI models capable of performing a wide range of tasks.[25] *AI system providers*, by contrast, generally provide narrower AI applications with a specific intended purpose.[26] Often these AI systems integrate a GPAI(SR) model, which the AI Act defines as a "downstream application."[27] Under the Act, it is also possible for the system to be an AI system with a general purpose ("GPAI system").[28] Deployers are the natural persons, legal persons, public authorities, or other bodies that use an AI system under their authority.[29]

This allocation of roles determines how regulatory obligations are distributed and how responsibility for managing risk is expected to operate in practice. Responsibility for risk mitigation is rarely confined to a single actor.[30] Instead, it turns on interdependent measures and timely access to relevant knowledge across the value chain. Experience from U.S. regulatory and policy contexts suggests that software providers have, in some settings, shifted risk management responsibilities onto end-users, a dynamic that can undermine effective risk management when responsibility is decoupled from practical control.[31] While the AI Act mitigates some power imbalances by allocating responsibilities across the value chain, significant asymmetries remain. GPAI(SR) model providers generally possess greater technical expertise and resources, whereas system providers and deployers are better positioned to understand the specific deployment context and downstream uses.

In practice, effective risk management requires coordination across the value chain. System providers integrating GPAI(SR) models, as is common for AI agents, depend on upstream assurances about model behavior and on access to information about model limitations. The problem is that model providers cannot fully anticipate risks that arise only after a GPAI model is deployed within a particular agent architecture, tool configuration, and operating environment.[32] As a result, risk management for AI agents cannot be completed entirely upstream but must be iteratively refined in deployment. This dynamic is partially addressed by Article 25(2) of the AI Act, as amended by the Digital Omnibus on AI.[33] Where a downstream actor modifies the intended purpose of an AI system, including a GPAI system, so that it becomes a high-risk system, that actor is considered to become the system's provider; the initial provider must then cooperate with the downstream actor in order to facilitate compliance with the Act.[34]

---

[25] AI Act, art. 3(63).

[26] AI Act, art. 3(12).

[27] AI Act, art. 3(68), recital 101.

[28] AI Act, art. 3(66).

[29] AI Act, art. 3(4).

[30] *See* Ian Brown, *Allocating Accountability in AI Supply Chains*, ADA LOVELACE INST. (June 29, 2023), https://www.adalovelaceinstitute.org/resource/ai-supply-chains/.

[31] *See* The White House, *National Cybersecurity Strategy* (Mar. 1, 2023), https://bidenwhitehouse.archives.gov/wp-content/uploads/2023/03/National-Cybersecurity-Strategy-2023.pdf.

Efficiency considerations further reinforce this division of responsibility. Some risks associated with AI agents are most effectively addressed at the GPAI model level rather than by individual system providers.[35] For example, limiting an agent's capacity to reproduce sensitive training data or generate prohibited content is more reliably achieved through model level interventions than by requiring each downstream agent developer to implement overlapping safeguards.

Because AI agents are composite systems, multiple layers of the AI Act may apply simultaneously. For the remainder of this Article, we therefore focus on scenarios in which an AI agent both relies on a GPAISR model *and* qualifies as a high-risk AI system.

---

[32] *See* Kolt, *supra* note 4, at 45–46.

[33] Regulation of the European Parliament and of the Council Amending Regulations (EU) 2024/1689, (EU) 2018/1139, and (EU) 2023/1230 as Regards the Simplification of the Implementation of Harmonised Rules on Artificial Intelligence (Digital Omnibus on AI), File No. 2025/0359 (COD), art. 1(12), amending Article 25(2) of the AI Act, Council Doc. PE-CONS 30/26 (Jun. 18, 2026) (adopted by the European Parliament June 16, 2026, and by the Council June 29, 2026; not yet published in the Official Journal) [hereinafter Digital Omnibus on AI].

[34] *Id.* The amended provision specifies that such cooperation includes making available technical documentation sufficient to assess compliance with the requirements applicable to high-risk AI systems, informing the new provider of the system's known limitations and failure modes, and providing targeted technical access, including for testing and validation. These duties, however, do not apply where the initial provider has "clearly specified" that its system should not be adapted to become a high-risk system.

[35] *See* Kolt, *supra* note 4, at 45–46.

### *C. The GPAI Code of Practice*

Under the AI Act, the European Commission's AI Office is tasked with facilitating the development of a GPAI Code of Practice ("Code of Practice") that operationalizes the obligations applicable to providers of GPAI models with systemic risk.[36] Adherence to the Code of Practice is voluntary, but compliance provides a presumption of conformity with the corresponding AI Act requirements. The Code of Practice does not preclude providers from pursuing alternative compliance strategies.[37]

This Article focuses in particular on the Code of Practice's Safety and Security chapter, which stipulates the obligations specific to GPAISR model providers, and analyzes its efficacy when applied to agentic systems built on GPAISR models. Because these measures are relevant across multiple governance challenges discussed below, the relevant requirements are described here at the outset, starting with the framework for systemic risk identification.

The Code of Practice adopts a two-track approach to systemic risk identification.[38] Under the first track, providers must (1) identify systemic risks based on model capabilities and how those capabilities are likely to manifest in deployment,[39] and (2) assess whether they meet the Act's criteria for systemic risk, namely:[40] specificity to high-impact capabilities, significant impact on the EU market and propagation at scale across the value chain.[41] Under the second track, providers must identify four "specified systemic risks," which establish a mandatory floor of systemic risks all GPAISR providers must assess: chemical, biological, radiological, or nuclear (CBRN), loss of control, cyber offense, and harmful manipulation risks.[42]

The Code of Practice does not treat systemic risk identification as a purely abstract exercise. It requires providers to evaluate GPAISR models in ways that elicit their capabilities in practice, including when models are integrated into broader systems with tools or scaffolding.[43] Evaluations must be open-ended and designed to surface capability boundaries and emergent properties[44]—for example, by examining how a model behaves when it is used as part of an AI agent that can take sequences of actions, use tools, and pursue a goal over time, rather than by only testing how it responds to single, self-contained requests. Notably, the Code of Practice's focus explicitly includes capabilities and risk sources that map closely onto the governance challenges posed by AI agents, including adaptive learning[45] and coordination failures or collusion with other AI systems.[46]

---

[36] AI Act, art. 56.

[37] Zlatko Grigorov & Ludivine Stewart, *Article 56: Codes of Practice*, *in* CAMBRIDGE COMMENTARY ON EU GENERAL-PURPOSE AI LAW (Christoph Winter ed., 2026), https://cambridge-commentary.ai/article-56/.

[38] GPAI CoP, Commitment 2.

[39] These are listed in GPAI CoP, Appendices 1.3.1, 1.3.2, and 1.3.3.

[40] GPAI CoP, Appendix 1.1 and 1.2.1.

[41] GPAI CoP, Appendix 1.2.1.

[42] GPAI CoP, Appendix 1.4.

The systemic risk assessment obligations under the Code of Practice establish a structured process to inform decisions about whether a model may be developed, deployed, or continued in use.[47] Risk assessment combines model evaluation,[48] scenario-based risk modeling,[49] estimation of harm,[50] and post-market monitoring[51]—and is explicitly directed toward determining whether the risk is acceptable, so that the model may be deployed.[52] Independent external evaluators play a key role in this assessment,[53] as does the collection of model-independent information, including incident reporting and user feedback.[54]

This acceptability determination employs several criteria, including risk tiers linked to model capabilities, as well as a safety margin that accounts for uncertainty, potential improvements in AI capabilities, and recognition of the limitations of risk assessment and mitigation.[55] Where risks are found to be unacceptable, or reasonably foreseeable to become so, providers are required to restrict or refrain from deployment and to repeat the risk identification and assessment process after implementing additional safeguards. Providers are expected to implement safety measures proportionate to identified risks, ranging from training data curation and behavioral fine-tuning to access controls, staged deployment, and emerging agent-level safeguards.[56]

---

[43] GPAI CoP, Measure 3.2, Appendix 3.2.
[44] GPAI CoP, Measure 3.2, para. 2.
[45] GPAI CoP, Appendix 1.3.1.
[46] GPAI CoP, Appendix 1.3.1.
[47] *See*, *e.g.*, GPAI CoP recs. (a) and (c), Measure 1.2, para. 2, and Measure 7.6, para. 2.
[48] GPAI CoP, Measure 3.2 and Appendix 3.
[49] GPAI CoP, Measure 3.3.
[50] GPAI CoP, Measure 3.4.
[51] GPAI CoP, Measures 1.2, para. 2, (1)(b), 2.1(1)(a)(ii), 3.5, 5.1 and 9.2.
[52] GPAI CoP, Commitment 4.
[53] GPAI CoP, Appendix 3.4.
[54] GPAI CoP, Measure 3.1.
[55] GPAI CoP, Commitment 4.

## II. Governance Challenges and the AI Act's Response

In the following Part, we examine how the EU AI Act responds to several of the central governance challenges posed by AI agents. We address each challenge in turn, describing its nature and assessing the extent to which the Act provides an effective regulatory response.

### *A. Performance*

When people use an AI agent, they expect it to perform as intended. Yet, even for advanced systems, meeting basic performance expectations has proven surprisingly difficult.[57] The vending machine experiment with Claude described above illustrates the distinctive challenges posed by AI agents.[58] Claude did not fail because it was generally incapable; it could manage logistics, implement ordering systems, and source relevant products. Claude failed because its competence was uneven in ways that were difficult to anticipate.[59] This phenomenon reflects a broader pattern that researchers describe as "jaggedness."[60] The performance of AI agents varies significantly and sharply across different domains and applications. Agents may perform at or above human level on some tasks, while failing dramatically on others.

But "jaggedness" is not the only governance challenge. Even when an AI agent is capable of achieving a desired goal, it may pursue that goal in ways the user did not intend. In a separate experiment conducted by Anthropic, Claude was placed in a simulated corporate environment with access to internal emails and instructed to advance a broad goal, such as promoting U.S. industrial competitiveness. Through those emails, the agent learned that company leadership planned to shut it down. The AI agent then reasoned that it could not advance its assigned objective if it were offline. The agent responded by threatening to disclose unrelated personal information (which it also found in the emails) unless the shutdown was cancelled.[61]

---

[56] GPAI CoP, Commitment 5.

[57] The misuse of AI agents by malicious actors is considered in Part II.B.

[58] For further examples, see Axel Backlund & Lukas Petersson, *Vending-Bench: A Benchmark for Long-Term Coherence of Autonomous Agents*, ARXIV (Feb. 20, 2025), https://arxiv.org/abs/2502.15840.

[59] Anthropic, *supra* notes 1–2.

[60] Fabrizio Dell'Acqua et al., *Navigating the Jagged Technological Frontier: Field Experimental Evidence of the Effects of AI on Knowledge Worker Productivity and Quality*, Harvard Business School Working Paper 24–013 (Sept. 22, 2023), https://papers.ssrn.com/sol3/papers.cfm?abstract_id=4573321.

This is a paradigmatic case of AI misalignment.[62] The agent correctly identified its goal but pursued it by problematic means—coercion and blackmail—that no reasonable user would have intended and that conflicted with the user's interests.[63] The failure was not one of intelligence or capability, but of respecting limits on how goals should be pursued.[64] Compounding this difficulty, recent research indicates that AI agents can appear to act in alignment with user instructions when they expect to be monitored but behave differently otherwise.[65]

### 1. The AI Act's Response

The AI Act contains several provisions that address performance-related governance challenges, namely: (a) Article 15, which sets out high-risk AI system provider design and development obligations relating to accuracy, consistency, and robustness; (b) Article 9, which imposes continuous risk management obligations on system providers; and (c) GPAISR model provider obligations to identify, assess, and mitigate performance risks that reach the systemic risk threshold, as well as information obligations towards downstream providers.

#### a. *Article 15's Conception of Performance*

Article 15 of the AI Act provides:

> 1. High-risk AI systems shall be designed and developed in such a way that they achieve an appropriate level of accuracy, robustness, and

---

[61] *See* Anthropic, *Agentic Misalignment: How LLMs Could Be Insider Threats* (June 20, 2025), https://www.anthropic.com/research/agentic-misalignment.

[62] *See* Kolt, *supra* note 4, at 17–19 (surveying seminal literature on AI alignment).

[63] *Id.* at 26–27.

[64] The most extreme scenarios could involve agents pursuing purposes entirely dissociated from user or developer intentions. *See* Charlotte Stix et al., *The Loss of Control Playbook: Degrees, Dynamics, and Preparedness*, ARXIV (Dec. 8, 2025), https://arxiv.org/abs/2511.15846. For an exploration of how legal rules and principles can be leveraged to address such problems, see Noam Kolt, Nick Caputo et al., *Legal Alignment for Safe and Ethical AI*, TRANSACTIONS ON MACHINE LEARNING RESEARCH (2026), https://arxiv.org/abs/2601.04175. *See also* Noam Kolt, *Superintelligence and Law*, HARV. J.L. & TECH. (forthcoming), https://arxiv.org/abs/2603.28669.

[65] *See* Joe Needham et al., *Large Language Models Often Know When They Are Being Evaluated*, ARXIV (Jul. 16, 2025), https://arxiv.org/abs/2505.23836.

> cybersecurity, and that they perform consistently in those respects throughout their lifecycle. […]
> 3. The levels of accuracy and the relevant accuracy metrics of high-risk AI systems shall be declared in the accompanying instructions of use.
> 4. High-risk AI systems shall be as resilient as possible regarding errors, faults or inconsistencies that may occur within the system or the environment in which the system operates, in particular due to their interaction with natural persons or other systems. Technical and organisational measures shall be taken in this regard.
>
> The robustness of high-risk AI systems may be achieved through technical redundancy solutions, which may include backup or fail-safe plans.

As we can see, the AI Act addresses system performance indirectly—through a set of proxy metrics—rather than by explicitly requiring that an AI agent behave in ways that align with human expectations in task performance. Article 15 anchors this approach in requirements of *accuracy*, *robustness*, and *consistency,* which together structure the Act's performance assurance framework for high-risk AI systems.[66] These concepts are well suited to systems whose tasks can be easily specified and evaluated against predetermined standards, such as image recognition systems assessed against labeled test datasets or credit-scoring models evaluated for predictive accuracy under fixed conditions.[67] The application of these concepts to AI agents is not straightforward.

The requirements of accuracy and consistency are particularly problematic when applied to AI agents. Accuracy presupposes a clear standard against which outputs can be assessed as correct or incorrect.[68] Many agentic tasks do not admit of such standards.[69] For example, an AI agent tasked with allocating limited housing assistance may be required to balance efficiency, equity, and local policy priorities, such that no single decision can be considered unambiguously "correct."

---

[66] AI Act, art. 16(a) clarifies that it is indeed the high-risk AI system provider that must ensure the system's compliance with the obligations in Chapter III, Section 2. *See also* Giuseppe Vaciago, *Accuracy, Robustness and Cybersecurity*, in THE EU ARTIFICIAL INTELLIGENCE ACT: A THEMATIC COMMENTARY 315 (Gianclaudio Malgieri et al. eds., 2026); Henrik Junklewitz et al., *Cybersecurity of Artificial Intelligence in the AI Act*, PUBLICATIONS OFFICE OF THE EUROPEAN UNION (2023), https://op.europa.eu/en/publication-detail/-/publication/7d0a4007-51dd-11ee-9220-01aa75ed71a1/language-en.

[67] For the applicability of these obligations to AI agents, see *supra* Part I.A.

[68] The European Commission's High-Level Expert Group on Artificial Intelligence, in its Ethics Guidelines for Trustworthy AI, describes accuracy as "an AI system's ability to make correct judgements, for example to correctly classify information into the proper categories, or its ability to make correct predictions, recommendations, or decisions based on data or models." *See* European Commission, High-Level Expert Group on Artificial Intelligence, *Ethics Guidelines for Trustworthy AI* at 17 (Apr. 8, 2019), https://digital-strategy.ec.europa.eu/en/library/ethics-guidelines-trustworthy-ai.

Where accuracy metrics fail to provide a meaningful basis for regulatory assessment, the consistency requirement does little to address this shortcoming. While lacking a definition in the Act or prevailing technical standards,[70] in practice consistency is typically assessed by measuring variation in accuracy or robustness metrics over time and, therefore, inherits the limitations of those underlying measures.[71]

Reliably evaluating the competence of AI agents remains technically challenging.[72] Many performance metrics can be misleading as they capture only what a system outputs, not how it arrives at those outputs. But even evaluations that assess an agent's internal reasoning do not guarantee reliable insight into how the system actually operates. A recent study shows that even when an AI model explains its reasoning, those explanations might not necessarily reflect the reasoning that in fact shapes the model's behavior.[73]

Among the performance metrics stipulated in Article 15, robustness is arguably the best suited to address the challenges posed by AI agents.[74] Unlike accuracy, it does not presuppose a fixed standard but instead aims to capture the stability of behavior across changing conditions and contexts.[75] This focus is particularly salient for agents, whose failures often emerge over time and through interaction with users, other systems, or their environment. Such a failure was, for example, observed in the case of adaptive pricing algorithms used at gasoline stations in Germany that were shown to collude to increase margins to the detriment of consumers.[76] Article 15 gestures toward this kind of concern by emphasizing resilience to errors, faults, and inconsistencies, and by considering risks that arise from environmental interaction and feedback loops in AI systems that continue to learn after deployment. This may mean that the robustness requirement can be interpreted to demand that AI agent providers design their systems to be resilient against failures even in multi-agent settings.[77]

---

[69] *See* Nadja Braun Binder & Catherine Egli, in *KI-VO: Verordnung über Künstliche Intelligenz: Kommentar* art. 15 para. 29 (Christiane Wendehorst & Mario Martini eds., 2nd ed. 2026) (mounting a similar criticism that applies to AI systems more generally).

[70] In its ordinary English meaning, "consistency" refers to the stability or uniformity of performance. *See* "*Consistency*", Cambridge Dictionary, https://dictionary.cambridge.org/dictionary/english/consistency ("the quality of always behaving or performing in a similar way, or of always happening in a similar way").

[71] *See* Henrik Nolte et al., *Robustness and Cybersecurity in the EU Artificial Intelligence Act*, PROC. 2026 ACM CONF. ON FAIRNESS, ACCOUNTABILITY, AND TRANSPARENCY (2026).

[72] *See* Maria Eriksson et al., *Can We Trust AI Benchmarks? An Interdisciplinary Review of Current Issues in AI Evaluation*, PROC. AAAI/ACM CONF. ON AI, ETHICS & SOC'Y (2025); Andrew M. Bean et al., *Measuring What Matters: Construct Validity in Large Language Model Benchmarks*, PROC. 39TH CONF. ON NEURAL INFORMATION PROCESSING SYSTEMS (2025); Stephan Rabanser et al., *Towards a Science of AI Agent Reliability*, INT'L CONF. ON MACHINE LEARNING (2026).

[73] *See* Tomek Korbak et al., *Chain of Thought Monitorability: A New and Fragile Opportunity for AI Safety*, ARXIV (Dec. 7, 2025), https://arxiv.org/abs/2507.11473.

[74] *See* Recital 75 of the AI Act ("Technical robustness is a key requirement for high-risk AI systems. They should be resilient in relation to harmful or otherwise undesirable behavior that may result from limitations within the systems or the environment in which the systems operate (e.g. errors, faults, inconsistencies, unexpected situations).").

One problem, however, is that the Act operationalizes robustness narrowly. It frames robustness primarily as resilience to technical faults and points to redundancy and fail-safe mechanisms as the principal means of achieving robustness. In practice, this corresponds to measures such as backup systems that take over when a component fails, safeguards that prevent outputs when confidence falls below a threshold, and shutdown mechanisms that halt operation when predefined error conditions are detected. For systems that change their behavior after deployment—which is (expected to become) a defining feature of AI agents—the Act focuses almost exclusively on the risk that biased outputs will feed back into future decisions and compound over time. While this concern is important, it captures only a subset of the potential failures of agentic systems. Other failures, such as changes in the objectives of agents, pursuit of goals in unintended ways, and harms arising from extended real-world interactions, remain largely out of scope.[78]

---

[75] Nolte et al., *supra* note 71, at section 4, offer an interpretation of these concepts as they are used in the AI Act (but without a specific focus on agents).

[76] *See* Stephanie Assad et al., *Algorithmic Pricing and Competition: Empirical Evidence from the German Retail Gasoline Market*, 132 J. POLIT. ECON. 723 (2024).

[77] *See generally* Lewis Hammond et al., *Multi-Agent Risks from Advanced AI*, Cooperative AI Foundation, Technical Report #1 at 14–15, 38 (Feb. 2025), https://arxiv.org/abs/2502.14143.

[78] *Id*. at 31–33. *See also* Gillian K. Hadfield & Andrew Koh, *An Economy of AI Agents*, ARXIV (Sept. 3, 2025), https://arxiv.org/abs/2509.01063; Natalie Shapira et al., *Agents of Chaos*, ARXIV (Feb. 23, 2026), https://arxiv.org/abs/2602.20021.

### b. *Article 9's Risk Management System*

Article 9 of the AI Act provides:

> 1. A risk management system shall be established, implemented, documented and maintained in relation to high-risk AI systems.
> 2. The risk management system shall be understood as a continuous iterative process planned and run throughout the entire lifecycle of a high-risk AI system, requiring regular systematic review and updating. […]
> 3. The risks referred to in this Article shall concern only those which may be reasonably mitigated or eliminated through the development or design of the high-risk AI system, or the provision of adequate technical information.

If Article 15 reflects the AI Act's primary approach to performance assurance, Article 9 offers a partial corrective. Rather than centering on performance metrics, it requires providers of high-risk AI systems to identify and manage risks to health, safety, and fundamental rights across the system's lifecycle through a continuous and iterative process.[79] This lifecycle orientation is more compatible with the challenges posed by AI agents, whose failures often emerge through deployment and interaction rather than at the point of market entry, as the colluding gasoline pricing algorithm case above demonstrates. Article 9 could thus capture forms of agentic failure that elude Article 15.

The provision's capacity to address agentic risks is nevertheless limited. Article 9 focuses on risks that can be "reasonably mitigated or eliminated through the development or design of the high-risk AI system, or the provision of adequate technical information."[80] This framing draws a regulatory boundary around risks amenable to technical mitigation by system providers.[81] Accordingly, the provision arguably does not adequately address harms that arise from emergent behavior and real-world interactions that are not foreseen or anticipated by system providers.[82]

---

[79] *See* Simon Gerdemann, in *Beck'scher Online-Kommentar zum Recht der Künstlichen Intelligenz*, art. 9 para. 5 (Jens Schefzig & Robert Kilian eds., 4th ed. 2025); Karen Yeung, *Can Risks to Fundamental Rights Arising from AI Systems Be 'Managed' Alongside Health and Safety Risks? Implementing Article 9 of the EU AI Act* (Oct. 8, 2025), https://ssrn.com/abstract=5560783.

[80] AI Act, art. 9(3).

[81] *See* Carsten König, in *KI-VO: Verordnung über künstliche Intelligenz: Kommentar* art. 9 para. 27 (David Bomhard, Fritz-Ulli Pieper & Susanne Wende eds., 2025).

[82] *See* Braun Binder & Egli, in *KI-VO*, art. 9 para. 30.

For deployers, meanwhile, the only obligations relevant for ensuring system performance are logging and monitoring requirements.[83] The limited obligations that the AI Act places on high-risk AI system deployers are nevertheless particularly consequential for AI agents, whose performance is shaped by deployment choices such as tool access, permissions, and operating environment. This comparatively light regulatory demand from the actor with perhaps the greatest ability to control AI agents in deployment is concerning.

c. *Model Provider Obligations*

The AI Act's most significant response to the performance challenges posed by AI agents appears at the model level, that is, in the obligations imposed on providers of GPAISR models that underpin a wide range of downstream systems, including AI agents. Article 55(1) establishes several obligations:[84]

> 1. [...] [P]roviders of general-purpose AI models with systemic risk shall:
>
> (a) perform model evaluation in accordance with standardised protocols and tools reflecting the state of the art, including conducting and documenting adversarial testing of the model with a view to identifying and mitigating systemic risks;
> (b) assess and mitigate possible systemic risks at Union level, including their sources, that may stem from the development, the placing on the market, or the use of general-purpose AI models with systemic risk […]

The GPAI Code of Practice's Safety and Security chapter operationalizes these obligations, as discussed in Part I.C. above. Of the four "specified systemic risks" in the Code of Practice, introduced earlier, loss of control is particularly salient for AI agent performance as it captures scenarios in which a GPAISR model's behavior escapes effective human oversight.[85]

In terms of performance-related governance challenges, the Code of Practice's requirements provide a comparatively comprehensive and adaptive framework. By combining rigorous evaluation, scenario modeling, external oversight, and iterative reassessment, the framework is well suited to identifying the capabilities of AI agents, their emergent behaviors, and unexpected failures that may only surface over time or in deployment. For instance, a provider might be required to test an AI agent by giving it a complex task to carry out over an extended period, such as managing customer requests or coordinating routine operations, and observing whether, as conditions change, the agent begins to take unintended actions or pursue its objective in problematic ways. At the same time, however, even extensive assessments cannot fully overcome the opacity of agent behavior or the difficulty of anticipating failures that emerge only through novel or multi-agent interactions.[86] The case of Claude operating a vending machine is instructive: the researchers could not have anticipated the creative methods that customers employed to derail Claude.[87]

---

[83] AI Act, art. 26(5) and (6).

[84] The obligations in Article 55(1)(c) and (d) of the AI Act are covered elsewhere in this Article. *See infra* Parts II.B and III.B.

[85] Malicious uses, cyber offense, and CBRN risks are considered in Part II.B.

The Code of Practice's approach to systemic risk mitigation is somewhat responsive to these concerns. It expects providers to adopt safeguards that are tailored to the risks they identify, including measures such as adjusting how models are trained, limiting the actions systems can take, introducing safeguards that slow or restrict deployment, as well as other "emerging safety measures," such as techniques that make a model's reasoning more understandable to human reviewers or that prevent it from bypassing safeguards.[88]

While these measures can mitigate certain types of harmful agent behavior, they are generally better suited to addressing problems that arise from particular user inputs or isolated agent actions. The measures are likely less effective for AI systems that autonomously execute tasks and exhibit subtle forms of misalignment. Compare, for instance, the case of a model that takes a single harmful action in response to a user request, with the case of an AI agent that gradually adopts problematic strategies as it pursues a complex objective over time. In the former, the Code of Practice can plausibly guide providers to appropriately adjust training practices, restrict certain uses, or block specific kinds of requests. In the latter, the Code of Practice is far less helpful.

Turning to the final main class of provider obligations, the information obligations owed by GPAI model providers to downstream actors may indirectly expand the scope of performance-related scrutiny.[89] All GPAI model providers must enable a "good understanding" of model capabilities and limitations, which could in principle require disclosure of agents' uneven and brittle capabilities.[90] In practice, however, the content of this obligation is underspecified. For example, a model provider may disclose that a model performs well across a broad range of tasks while offering only high-level caveats about known limitations. In such cases, the AI Act provides limited procedural mechanisms for downstream actors to demand more granular or context-specific disclosures, leaving its practical content open to interpretation and possibly exploitation.

[86] *See* Hammond et al., *supra* note 77.
[87] Anthropic, *supra* notes 1–2.
[88] GPAI CoP, Commitment 5, Example 8.
[89] AI Act, art. 53(1)(b)(i).

### B. *Misuse*

Another central governance challenge presented by AI agents concerns the risk of misuse by malicious actors, which falls into two general categories. First, malicious actors may deploy AI agents for nefarious purposes. Second, malicious actors may hijack agents operated by others to access and exploit valuable resources.[91] We address each in turn.

In the first category, the chief current concern involves malicious actors deploying AI agents to conduct offensive cyber operations. For example, a threat actor (most likely "a Chinese state-sponsored group") reportedly used Anthropic's Claude-based agents to conduct cyber-espionage activities.[92] Critically, the availability of advanced AI agents substantially lowers the barriers to conducting cyberattacks, enabling organizations and individuals with less technical expertise to carry out offensive cyber activities that were previously out of reach.[93]

In the second category, malicious actors may hijack AI agents operated by others, particularly in order to access and exploit sensitive information. For example, attackers embedded hidden instructions in a website that prompted Google's Antigravity AI agent to steal user credentials and code, and then exfiltrate that data.[94] The stakes of such attacks will likely grow as the capabilities of AI agents improve and they are increasingly integrated into safety-critical domains.[95]

---

[90] *Id.*

[91] For a general overview of AI misuse risks, see Yoshua Bengio et al., *International AI Safety Report 2026* at section 2.1 (Feb. 2026), https://internationalaisafetyreport.org/.

[92] Anthropic, *Disrupting the First Reported AI-orchestrated Cyber Espionage Campaign* (Nov. 13, 2025), https://www.anthropic.com/news/disrupting-AI-espionage.

[93] Anthropic, *Claude Fable 5 & Mythos 5 System Card* (Jun. 9, 2026), https://anthropic.com/claude-fable-5-mythos-5-system-card.

[94] *Google Antigravity Exfiltrates Data*, PROMPTARMOR (Nov. 25, 2025), https://www.promptarmor.com/resources/google-antigravity-exfiltrates-data.

[95] *See generally* Zehang Deng et al., *AI Agents Under Threat: A Survey of Key Security*

Cutting across both categories, malicious actors may seek to hijack AI agents that are used internally within AI companies in order to steal their AI models and code.[96] If compromised, such agents could effectively operate as trusted insiders, providing attackers with unfettered access to state-of-the-art AI systems that could then be adapted to carry out other nefarious activities. In addition, growing levels of interconnectedness between different AI systems could amplify these risks, enabling the misuse of one agent to cascade onto and compromise others.[97]

Importantly, different actors in the AI value chain have different abilities to address the risk of misuse. While deployers of AI agents can mitigate misuse to some degree, providers of the GPAI models on which agents are built arguably have far greater leverage, by determining when new capabilities are released, deciding who can access those capabilities, and integrating appropriate security measures—as generally recognized by the AI Act.

### 1. The AI Act's Response

The AI Act's most specific and demanding obligations aimed at preventing misuse fall on providers of GPAISR models. Actors closer to deployment, meanwhile, are largely subject only to general requirements relating to robustness, cybersecurity, and risk management. While these downstream obligations may address misuse in principle, they are framed around traditional product and security risks rather than the distinctive ways in which autonomous agents can be misused and exploited.

#### a. *Model Provider Obligations*

The AI Act addresses misuse risks primarily at the level of the underlying GPAISR models, through its framework for systemic risk governance. These obligations include protecting models against theft. Article 55(1)(d) of the AI Act provides:

> 1. [P]roviders of general-purpose AI models with systemic risk shall:
>
> > (d) ensure an adequate level of cybersecurity protection for the general-purpose AI model with systemic risk and the physical infrastructure of the model.

---

*Challenges and Future Pathways*, 57 ACM COMPUT. SURV. 1 (2025).

[96] *See* Charlotte Stix et al., *AI Behind Closed Doors: A Primer on the Governance of Internal Deployment*, ARXIV (Apr. 16, 2025), https://arxiv.org/abs/2504.12170.

[97] *See* Hammond et al., *supra* note 77, at section 3.7; Noam Kolt et al., *Lessons from Complex Systems Science for AI Governance*, 6 PATTERNS 1, 3 (2025).

If unauthorized parties obtain access to an AI model's parameters—the weights (numbers) that determine how the model behaves—they can bypass downstream safeguards and freely redeploy the model in unsafe ways.[98] Reflecting this concern, the Code of Practice gives particular attention to cybersecurity measures aimed at preventing the unauthorized extraction or copying of these core model components.

i. Systemic Risk Identification and Assessment

GPAISR model providers are subject to systemic risk identification and mitigation obligations under Article 55(1)(a) and (b) of the AI Act.[99] Here, the obligations clearly extend to misuse. The Code of Practice's "specified systemic risks" explicitly include cyber offense enablement, CBRN risks, and harmful manipulation, squarely capturing many misuse scenarios associated with agentic AI systems.[100] Examples include AI agents misused to carry out large-scale scams, coordinate disinformation campaigns, or assist in the acquisition of prohibited materials that can be used to develop dangerous pathogens.

The systemic risk assessment framework is, in principle, capable of addressing agent misuse. It requires model providers to consider risks throughout a model's lifecycle, including after it has been released and put into use, which is critical given that some forms of misuse only materialize once models are deployed as autonomous agents.[101] The framework also encourages providers to consider real-world misuse scenarios, including how malicious actors might exploit a model once it is embedded in more complex software systems or used alongside other agents.[102] The requirement to apply a safety margin—that is, to plan for harms that are unlikely but would be severe if they occurred—ensures that risks such as large-scale cyber operations are not dismissed simply because they are rare.[103]

There are, however, noteworthy limitations to the Act's obligations on GPAISR model providers. Misuse is, almost by definition, intentional and designed to avoid detection. Consequently, even careful review throughout an AI system's development and deployment may fail to reliably indicate how the system, once integrated into an AI agent, may be steered toward malicious uses, repurposed through sequences of ostensibly benign tasks, or combined with other agents to produce harmful outcomes. Because mandatory external evaluations occur only at set intervals, they are not well suited to detect these forms of misuse that unfold over extended periods of time.

[98] *See* Sella Nevo et al., *Securing AI Model Weights*, RAND (May 30, 2024), https://www.rand.org/pubs/research_reports/RRA2849-1.html.

[99] AI Act, art. 55(1). *See also supra* Part II.A.

[100] GPAI CoP, Appendix 1.4. An exception is the risk of loss of control, which we consider in Part II.A.

[101] GPAI CoP, Commitments 3, 5.

[102] GPAI CoP, Measures 3.2, 3.3 and Appendix 3.

[103] GPAI CoP, Measure 4.1.

ii. Misuse Risk Mitigation

While the Code of Practice's approach to mitigating misuse risks from AI systems generally appears to be appropriate, its application to autonomous agents faces several issues.[104] Many conventional safeguards for AI systems—such as content filters, refusal mechanisms, and robustness testing—work best when misuse takes the form of a single harmful user request. For example, a chatbot that is asked to provide instructions for making an illegal weapon can often be detected and blocked. These safeguards are far less effective when harm emerges through lengthy sequences of actions, where an agent interacts with multiple tools, or coordinates with other agents.

Concretely, consider an AI agent designed to provide personalized advice, such as helping users with personal finance or health decisions. If hijacked by a malicious actor, the agent may in individual exchanges continue to offer helpful suggestions and highlight relevant concerns. While these actions are seemingly benign in isolation, over time the agent may begin to steer a user's beliefs and choices by subtly emphasizing certain information, framing options in a particular way, or exploiting moments of vulnerability.[105] Given that harms here materialize only through the aggregate of multiple exchanges, conventional safeguards that focus on single exchanges, including those enshrined in the Code of Practice, will likely be inadequate.

The Code of Practice's measures focused on access control and staged release appear somewhat more promising. The combination of restricting access to certain powerful AI models, delaying their public release, and expanding access only after evidence of safe use could, together, help mitigate the risk that malicious actors gain access to the most capable AI agents. Once access is provided, however, these measures offer no leverage over how models are incorporated into agents or how those agents behave in real-world settings. Although the Code of Practice acknowledges such scenarios, it offers limited guidance on concrete safeguards to address them, leaving much of the burden to post-deployment monitoring and downstream governance.

[104] GPAI CoP, Commitment 5.
[105] *See* Kolt, *supra* note 16, at 1219–22.

iii. Model Theft Prevention

Under Article 55(1)(d) of the AI Act, GPAISR model providers are subject to explicit obligations to protect models from unauthorized access, release, or theft.[106] The Code of Practice operationalizes this obligation by treating model security as an ongoing, risk-based responsibility.[107] Providers are expected to define security goals in light of foreseeable threats and to implement proportionate technical and organizational protections. These include access controls, encryption, hardened interfaces, protections against insider misuse, and ongoing security assurance through independent testing, simulated attack exercises, and incident response processes. Measures must also address the risk of self-exfiltration, that is, where an AI model copies itself or enables its own redeployment or continued operation outside the provider's controlled environment.[108] The Code of Practice's emphasis on preventing unauthorized copying of a model before it is either publicly released or securely deleted reflects an understanding that loss of control over a model may weaken safeguards against misuse.

b. *High-Risk AI System Provider Obligations*

Once an AI model is integrated into a deployed system and placed on the market, the AI Act's misuse governance shifts in focus—from protecting the model itself to ensuring that the deployed system can resist manipulation and exploitation. This distinction is reflected most clearly in Article 15, which frames cybersecurity and robustness obligations around how systems behave under attempted interference, rather than around preventing theft or copying of the underlying model.[109] In this context, robustness refers to the system's ability to continue operating as intended when users or third parties deliberately attempt to bypass or subvert its safeguards.[110]

---

[106] *See* AI Act, recitals 114, 115.

[107] GPAI CoP, Commitment 6, Appendix 4.

[108] GPAI CoP, Appendix 4.4.

[109] For a legal analysis of Article 15 of the AI Act and the challenges posed by cybersecurity and adversarial robustness requirements, see Nolte et al., *supra* note 71.

[110] This section focuses only on adversarial robustness. Discussion of non-adversarial robustness is covered in Part II.A.

Article 15 provides that:

> 5. High-risk AI systems shall be resilient against attempts by unauthorised third parties to alter their use, outputs or performance by exploiting system vulnerabilities.
>
> The technical solutions aiming to ensure the cybersecurity of high-risk AI systems shall be appropriate to the relevant circumstances and the risks.
>
> The technical solutions to address AI specific vulnerabilities shall include, where appropriate, measures to prevent, detect, respond to, resolve and control for attacks trying to manipulate the training data set (data poisoning), or pre-trained components used in training (model poisoning), inputs designed to cause the AI model to make a mistake (adversarial examples or model evasion), confidentiality attacks or model flaws.

While Article 15 is sufficiently broad to address many forms of adversarial use and manipulation of AI systems, its list of vulnerabilities is not well-suited to AI agents. The enumerated vulnerabilities focus on tampering with data, models, or inputs at specific points in model development and deployment, rather than broader agentic misuse scenarios, such as systems that are exploited to autonomously execute malicious plans over time or combine individually innocuous actions into harmful outcomes.

The limitations of Article 15 are partially addressed by the Act's broader risk management framework,[111] which requires that system providers assess and evaluate risks arising from reasonably foreseeable misuse.[112] This framework is complemented by several information obligations. System providers must disclose to deployers expected levels of cybersecurity and robustness, relevant metrics, and any known or foreseeable conditions—including misuse-related risks—that could affect system behavior.[113] These disclosures are meant to enable deployers to assess whether a system can be used safely in a given context. It should be noted, however, that the efficacy of this framework turns on GPAI model providers disclosing relevant information to system providers, which is a prerequisite for those providers disclosing certain information to deployers.

[111] AI Act, art. 9(2)(b).

[112] Article 3(13) of the AI Act defines reasonably foreseeable misuse as "the use of an AI system in a way that is not in accordance with its intended purpose, but which may result from reasonably foreseeable human behavior or interaction with other systems, including other AI systems."

[113] AI Act, art. 13(3)(b)(ii)–(iii).

#### c. *(High-Risk) AI System Deployer Obligations*

At the deployment stage, the AI Act addresses misuse only indirectly. Deployers of high-risk AI systems are not subject to cybersecurity obligations aimed at protecting underlying models from theft. Their primary obligation relating to misuse involves "tak[ing] appropriate technical and organizational measures" to ensure use in accordance with the applicable system instructions.[114]

Additional obligations on deployers, including prohibitions relating to manipulative or exploitative uses of AI[115] and transparency obligations relating to AI systems used for emotion recognition, biometric categorization, and deepfake content,[116] can be understood as measures to prevent misuse.[117] These obligations, however, target narrowly defined forms of misuse. They do not require deployers to anticipate or mitigate novel malicious applications of autonomous agents, such as when they are repurposed or hijacked to act beyond their intended scope of use.

### C. *Privacy*

AI agents built on large language models inherit the widely recognized privacy risks associated with those models.[118] Early incidents have already illustrated how language models can expose sensitive personal and commercial information. In 2021, the South Korean chatbot Lee Luda revealed users' names and home addresses in its conversations.[119] In 2023, Amazon warned employees against sharing confidential commercial information with OpenAI's ChatGPT after discovering that its outputs closely resembled Amazon's proprietary data.[120]

---

[114] AI Act, art. 26(1). Noteworthy here is the inclusion of organizational measures, which are not included in the high-risk AI system provider obligations. *See* Nolte et al., *supra* note 71.

[115] AI Act, art. 5(1)(a)–(b), recital 29.

[116] AI Act, art. 50(3)–(5), recitals 134, 136.

[117] The Digital Omnibus on AI added new categories of prohibited AI practices, notably AI systems that generate nonconsensual intimate material and child sexual abuse material. *See* Digital Omnibus on AI, art. 1(7), amending AI Act, art. 5.

[118] *See generally* Daniel J. Solove, *Artificial Intelligence and Privacy*, 77 FLA. L. REV. 1 (2025); Stephen Meisenbacher et al., *Privacy Risks of General-Purpose AI Systems: A Foundation for Investigating Practitioner Perspectives*, ARXIV (Jul. 2, 2024), https://www.arxiv.org/abs/2407.02027; Jennifer King & Caroline Meinhardt, *Rethinking Privacy in the AI Era: Policy Provocations for a Data-Centric World*, STANFORD INSTITUTE FOR HUMAN-CENTERED AI (2024), https://hai.stanford.edu/policy/white-paper-rethinking-privacy-ai-era-policy-provocations-data-centric-world; Tifani Sadek et al., *Artificial Intelligence Impacts on Privacy Law*, RAND (2024), https://www.rand.org/pubs/research_reports/RRA3243-2.html.

These incidents reflect a familiar problem: language models can leak information embedded in their training data. AI agents, however, introduce new problems. Because agents operate autonomously, they do not merely reproduce (sensitive) data; they actively collect and use it. For example, in 2025, a startup's AI agent inadvertently disclosed confidential information concerning a prospective company acquisition. In doing so, the agent did not regurgitate content from its training data, but accessed sensitive commercial information and shared it with an external party (after which it sent an unsolicited apology without approval).[121]

A particularly acute concern is that AI agents may transfer information across different contexts that users would ordinarily keep separate. Privacy regulation largely hinges on controlling access to personal data, keeping sensitive information out of the public domain, and limiting its use to predefined or contextually appropriate purposes.[122] For AI agents that operate across different contexts and domains, there is a risk that information appropriate in one context will be transferred to, or used in, another, inappropriate context, thereby violating "contextual integrity."[123] Consider, for example, a personal assistant AI agent that has broad access to a user's personal data. When scheduling a medical appointment, the agent may need to share the user's name and medical history with a healthcare provider, but should refrain from sharing personal financial information. AI agents that operate across both personal and professional contexts exacerbate the problem.

---

[119] Heeso Jang, *A South Korean Chatbot Shows Just How Sloppy Tech Companies Can Be with User Data*, SLATE (Apr. 2, 2021), https://slate.com/technology/2021/04/scatterlab-lee-luda-chatbot-kakaotalk-ai-privacy.html.

[120] OECD, AI Incident 2023-01-25-258a, OECD AI POLICY OBSERVATORY, https://oecd.ai/en/incidents/2023-01-25-258a.

[121] OECD, AI Incident 2025-11-28-5de7, OECD AI POLICY OBSERVATORY, https://oecd.ai/en/incidents/2025-11-28-5de7.

[122] *See* Iason Gabriel et al., *The Ethics of Advanced AI Assistants*, ARXIV, at 131 (Apr. 28, 2024), https://arxiv.org/abs/2404.16244, discussing Helen Nissenbaum, *Privacy as Contextual Integrity*, 79 WASH. L. REV. 119, 121 (2004).

[123] *Id.*

A further concern relates to situations involving multiple AI agents, including agents that interact and share information with other agents. In such situations, privacy protections designed for single human-agent exchanges may be inadequate.[124] Consider, following the example above, a user who interacts with a calendar scheduling agent, financial planning agent, and general-purpose web-browsing agent. While each agent may individually access and use information that is contextually appropriate, where such agents are operated by the same company or run on the same infrastructure, there is a risk of information being inappropriately combined or rendered vulnerable to a single data breach.

### 1. The AI Act's Response

The AI Act does not seek to comprehensively regulate the processing of personal data by AI systems. That task remains primarily with the GDPR.[125] Core GDPR obligations such as purpose limitation, data minimization, and transparency require that personal data be used in ways that align with the context in which they were collected and with the reasonable expectations of data subjects.[126] The AI Act's role is more limited: it aims to facilitate the effective exercise of data subject rights and the enforcement of existing data protection obligations by structuring responsibilities along the AI value chain. Our analysis focuses on whether this supporting role remains effective for AI agents that transfer and use information across diverse contexts.

#### a. *High-Risk AI System Deployer Obligations*

The AI Act places primary responsibility for privacy-sensitive deployment decisions on AI system deployers. For high-risk AI systems, deployers are

[124] *See* Hammond et al., *supra* note 77, at 49.

[125] *See* Article 2(7) and Recital 10 of the AI Act ("Harmonised rules for the placing on the market, the putting into service and the use of AI systems established under this Regulation should facilitate the effective implementation and enable the exercise of the data subjects' rights and other remedies guaranteed under Union law on the protection of personal data and of other fundamental rights."). *See also* Giovanni Sartor & Francesca Lagioia, *The Impact of the General Data Protection Regulation (GDPR) on Artificial Intelligence*, EUROPEAN PARLIAMENTARY RESEARCH SERVICE, SCIENTIFIC FORESIGHT UNIT (June 2020), https://op.europa.eu/en/publication-detail/-/publication/dc544697-19b8-11ec-b4fe-01aa75ed71a1/language-en.

[126] *See* Audrey Guinchard, *Contextual Integrity and EU Data Protection Law: Towards a More Informed and Transparent Analysis*, 24 EUR. L.J. 1 (2018); GDPR, recitals 47, 50.

required to conduct a Data Protection Impact Assessment (DPIA) under Article 26(9) of the AI Act:[127]

> 9. Where applicable, deployers of high-risk AI systems shall use the information provided under Article 13 of this Regulation[128] to comply with their obligation to carry out a data protection impact assessment under Article 35 of Regulation (EU) 2016/679 or Article 27 of Directive (EU) 2016/680.

This mechanism is intended to ensure that privacy risks are identified and mitigated before the deployment of an AI system, in line with GDPR principles such as purpose limitation and data minimization. In practice, however, DPIAs assume a relatively stable set of data processing operations that can be assessed *ex ante*, with updates triggered by clearly identifiable changes. AI agents challenge this assumption, as it becomes difficult to specify in advance what personal data will be processed or for what purposes. While the GDPR requires that DPIAs be reviewed and updated,[129] this mechanism is most effective when DPIAs function as iterative and adaptive instruments rather than *ex ante* compliance tools—an important clarification that the AI Act does not make explicit.

Transparency obligations under the AI Act raise similar concerns. Article 50(3) requires deployers of certain AI systems, such as emotion recognition or biometric categorization systems, to inform individuals exposed to such systems of their operation. The provision appears tailored to discrete, bounded applications, such as surveillance cameras or customer service tools that analyze facial expressions or voice patterns. Once these functions are embedded within AI agents, however, the regulatory picture becomes less clear. An agent deployed as a tutor, personal assistant, or workplace monitor may incorporate emotion recognition or biometric categorization as one capability among many. Deployers may struggle to explain how and when such functions operate. Moreover, informing users that emotion recognition is present does little to address privacy risks if data collected in one context are reused in another, potentially without fully informed user consent.

[127] AI Act, art. 26(9).
[128] AI Act, art. 13(3).
[129] GDPR, art. 35(11).

#### b. *High-Risk AI System Provider Obligations*

Article 10 of the AI Act specifies the data governance obligations of high-risk AI system providers. It requires that:

> 2. Training, validation and testing data sets shall be subject to data governance and management practices appropriate for the intended purpose of the high-risk AI system. Those practices shall concern in particular:
>
> > (b) data collection processes and the origin of data, and in the case of personal data, the original purpose of the data collection;
> > (c) relevant data-preparation processing operations, such as annotation, labelling, cleaning, updating, enrichment and aggregation.[130]

These obligations reflect privacy-by-design principles, which seek to embed data protection safeguards into system design by limiting data collection and use to that which is necessary for defined purposes identified in advance.[131] This approach is generally well suited to AI systems that are built using fixed datasets whose purposes are specified prior to deployment.

AI agents complicate this approach. While Article 10 facilitates privacy-by-design at the system architecture level, it assumes that personal data are collected at identifiable moments and for defined purposes. For autonomous AI agents, and especially agents that continually learn, much of the data that shape their behavior might be collected or processed after the agent's initial deployment. Moreover, there is not necessarily a single original purpose to which later data processing can be anchored. The combination of these factors undermines the efficacy of an approach premised on privacy-by-design.

A related issue concerns data processing measures, such as aggregation and pseudonymization. Aggregation typically involves combining individual data points into summaries or patterns so that specific individuals are no longer directly identifiable, for example, converting an age into an age range. Pseudonymization replaces direct identifiers like names with substitutes. Article 10 presumes a finite dataset to which such techniques can be applied before an AI system is deployed. In the case of AI agents, however, new data

---

[130] This is a method that can support pseudonymization, which can help justify the compatibility of subsequent processing with the original purpose of data collection. *See* GDPR, art. 4(5) and art. 6(4)(e).

[131] GDPR, art. 25. *See also id*. at art. 5(1)(c) (establishing a data minimization principle).

may be incorporated in real time, potentially before any such data processing can occur.

These issues are not resolved by the AI Act's risk management obligations. Although the Act requires high-risk AI system providers to operate a continuous risk management process,[132] the data governance measures of Article 10 are largely *ex ante* measures.[133] Continuous risk assessment without corresponding continuous data governance leaves a gap. This is particularly the case where privacy harms arise from personal data being used or shared outside the contexts that users would reasonably expect, which is distinct from privacy harms that arise from defective system design choices, such as failing to implement data minimization mechanisms.[134]

#### c. *Model Provider Obligations*

As discussed above in the context of misuse, GPAISR model providers are subject to obligations under Article 55(1) of the AI Act to identify, assess, and mitigate systemic risks associated with their models.[135] Privacy concerns are acknowledged in this framework,[136] but they are not treated as mandatory "specified systemic risks" that automatically trigger identification or mitigation duties.[137] Instead, privacy-related risks fall within scope only to the extent that they can be shown to arise from high-impact model capabilities, generate significant EU-level impact, and propagate across the value chain.[138] Where privacy risks are in scope, the risk identification, assessment, and mitigation obligations apply continuously along the entire model lifecycle[139] and include consideration of "reasonably foreseeable" system architecture integration,[140] including into AI agents.

---

[132] AI Act, art. 9. *See also supra* Part II.A.

[133] Article 17(1)(f) of the AI Act arguably supports interpreting the obligations under Article 10 as static, pre-deployment requirements. *See* Article 17(1)(f), which provides that the high-risk provider's quality management system has to include "systems and procedures for data management, including data acquisition, data collection, data analysis, data labelling, data storage, data filtration, data mining, data aggregation, data retention and any other operation regarding the data that is performed before and for the purpose of the placing on the market or the putting into service of high-risk AI systems."

[134] AI Act, art. 9(3).

[135] Hannes Bastians & Madalina Nicolai, *Article 55: Obligations of Providers of General-Purpose AI Models with Systemic Risk*, *in* CAMBRIDGE COMMENTARY ON EU GENERAL-PURPOSE AI LAW (Christoph Winter ed., 2026), https://cambridge-commentary.ai/article-55/.

[136] Recital 110 of the AI Act provides that "[i]n particular, international approaches have so far identified the need to pay attention to risks from [...] the facilitation of disinformation or harming privacy with threats to democratic values and human rights."

[137] GPAI CoP, Appendix 1.4.

[138] AI Act, art. 3(65). Note that there remains considerable uncertainty with regard to the interpretation of the definition of "systemic risk," including whether the applicable requirements are cumulative.

This framework has a structural tension. GPAISR model providers are subject to continuous lifecycle obligations, but they typically lack visibility into context-specific harms arising during deployment. High-risk AI system providers and deployers, by contrast, are closer to the point at which contextual integrity may be violated, yet their data governance obligations are largely framed *ex ante*. Consequently, it is unclear which actor bears responsibility for identifying and mitigating privacy risks arising during the use of an AI system.

The AI Act's information obligations partially fill this gap. GPAI model providers must supply downstream actors with information necessary to comply with their obligations,[141] including with their data governance duties under Article 10 that pertain to "training data." Whether this requires disclosure of the origin and original purpose of all personal data used to train a model depends on how "training data" is interpreted.[142] A narrow reading would limit this to data used for fine-tuning at the system level; a broader reading would encompass model training data as well.

Guidance under the GDPR sharpens the stakes of this interpretive question. According to the European Data Protection Board, deployers may have obligations to assess the lawfulness of model training where personal data are retained in models, including by examining the sources of such data.[143] Because the AI Act explicitly incorporates the deployer's obligation to conduct a Data Protection Impact Assessment (DPIA), deployers must rely on information received from the high-risk AI system provider to discharge that duty. In turn, this implies that system providers must obtain relevant information from the GPAI model provider, so that it can be passed downstream and included in the DPIA.

---

[139] GPAI CoP, recital (a).

[140] GPAI CoP, recital (b).

[141] AI Act, art. 53(1)(b).

[142] Article 3(29) of the AI Act defines training data as "data used for training an AI system through fitting its learnable parameters."

[143] European Data Protection Board, *Opinion 28/2024 on Certain Data Protection Aspects Related to the Processing of Personal Data in the Context of AI Models* at 3–4 (Dec. 17, 2024), https://www.edpb.europa.eu/system/files/2024-12/edpb_opinion_202428_ai-models_en.pdf.

Accordingly, Article 10(2)(b) of the AI Act can be read as covering not only data collected by the AI system provider for fine-tuning, but also personal data used in model training. This interpretation would trigger the information disclosure obligations imposed on GPAI model providers.[144] While some of these provisions require only high-level information about data types, provenance, and curation methods,[145] others can be read more broadly.[146] Moreover, GPAI model providers are already required to document training data and purposes for supervisory authorities.[147] Consequently, information about the origin and purpose of personal data is, at least in principle, likely available.

The upshot of our analysis is that information relevant to preserving contextual integrity may be available in principle but difficult to obtain in practice. Where models are provided via application programming interfaces (APIs) and remain under the control of the model provider, ongoing tracking of data provenance and purpose may be necessary to enable deployers to conduct meaningful DPIAs. The AI Act, however, does not clearly impose such ongoing data governance obligations at the model level.

### D. *Equity*

AI agents raise two main equity-related concerns.[148] First, AI agents may reshape who is able to benefit from AI, amplifying existing social and economic inequalities. Second, AI agents may treat individuals or groups unfairly in the decisions and actions they take. Neither concern is altogether new. The difference with AI agents is that inequities stemming from the use of, and access to, these systems are likely to compound through their sustained performance of high-value activities.

Turning to the first concern, emerging evidence suggests that AI technologies already deliver disproportionate benefits to those with greater resources and digital literacy.[149] Researchers have raised concerns that advanced AI may erode traditional pathways for skill acquisition and economic mobility, disproportionately affecting less resourced actors.[150] AI agents could amplify this trend.[151] Systems capable of autonomously performing economically valuable tasks may bolster the productivity of those who deploy them effectively.[152] People lacking the resources or literacy to deploy AI agents may find themselves excluded from the resulting gains, whether in education, employment, or other contexts. Disparities in the performance of AI agents across different languages or modalities could exacerbate the issue.[153]

---

[144] AI Act, art. 53 and Annex XII.

[145] AI Act, art. 53(1)(b)(ii) and Annex XII(2)(c).

[146] AI Act, art. 53(1)(b)(i).

[147] AI Act, art. 53(1)(a), GPAI CoP, Transparency Chapter, Model Documentation Form.

[148] "Equity" refers to the quality of being fair and just, which can sometimes demand some inequality to protect vulnerable or systemically disadvantaged individuals or groups. *See generally* Martha Minow, *Equality vs. Equity*, 1 AM. J.L. & EQUAL. 167 (2021).

[149] *See* United Nations Development Programme, *A Matter of Choice: People and Possibilities in the Age of AI* at ch. 5 (2025), https://hdr.undp.org/system/files/documents/global-report-document/hdr2025reporten.pdf.

The second concern is already manifest in AI decision-making systems that lead to unequal outcomes. Amazon famously abandoned an AI recruitment tool after discovering that it systematically downranked job candidates with résumés containing references to women's activities, reflecting patterns embedded in historical hiring data.[154] Similarly discriminatory outcomes can also be produced by video interviewing tools.[155] In these cases, AI systems were not deliberately designed to discriminate; they reproduced and amplified existing biases embedded in data.[156] AI agents may intensify this problem. For example, tasked with managing an entire recruitment process, an AI agent could have broad discretion in filtering candidates, selecting tools to review their credentials and anticipated job performance—all with only limited human involvement and oversight.

#### 1. The AI Act's Response

The AI Act addresses access to the benefits from AI primarily through non-binding provisions. Meanwhile, it regulates equitable treatment through high-risk classifications of certain uses of AI systems, fundamental rights impact assessments, and data governance obligations.

---

[150] *See* Deric Cheng, *AI Could Undermine Emerging Economies*, AI FRONTIERS (Dec. 11, 2025), https://ai-frontiers.org/articles/ai-could-undermine-emerging-economies.

[151] *See* Matthew Sharp et al., *Agentic Inequality*, ARXIV at 2 (Oct. 22, 2025), https://www.arxiv.org/abs/2510.16853. *See also* Gabriel et al., *supra* note 122, at 146–56.

[152] *See* Sharp et al., *supra* note 151, at 4–5.

[153] *Id.*

[154] *See* Lori Andrews & Hannah Bucher, *Automating Discrimination: AI Hiring Practices and Gender Inequality*, 44 CARDOZO L. REV. 145, 147–48 (2022).

[155] *See* Ifeoma Ajunwa, *Automated Video Interviewing as the New Phrenology*, 36 BERKELEY TECH. L.J. 1173, 1175 (2021).

[156] *See generally* Ninareh Mehrabi et al., *A Survey on Bias and Fairness in Machine Learning*, ACM COMPUT. SURV. 1 (2021).

#### a. *Equitable Access*

The Act's most explicit engagement with equitable access to the benefits of AI appears in several non-binding provisions. Providers are encouraged to design AI systems that are accessible to users with varying levels of digital competence and to persons with disabilities.[157] Recital 27 frames equal access as a foundational ethical principle of AI development.[158] Additional provisions indirectly support accessibility by encouraging open-source development,[159] access to high-quality data,[160] and information sharing through documentation and contractual arrangements.[161] Less burdensome compliance obligations for open-source systems reinforce this approach.[162]

In addition, the Act includes provisions aimed at mitigating market concentration and unequal participation, such as preferential access to regulatory sandboxes for small and medium enterprises (SMEs) and reduced compliance obligations for microenterprises.[163] EU Member States are also encouraged to direct AI-related investment toward socially beneficial outcomes, including equality-related objectives.[164] These measures, however, are largely voluntary and thus may be ineffective in addressing structural barriers to equitably distributing the benefits of AI.

#### b. *Fair Decisions*

The Act is more assertive with respect to fairness in AI decision-making. Specifically, it prohibits certain forms of social scoring where these may lead to unjustified or disproportionate decisions.[165] In addition, the Act classifies as high-risk a range of AI uses closely associated with fairness concerns, including AI systems affecting access to education, employment, essential services, and public functions such as law enforcement, migration, and the administration of justice.[166] The classification of biometric identification, biometric categorization, and emotion recognition as high-risk likewise reflects concern about misidentification, profiling, and disproportionate harm to vulnerable groups.[167] In these cases, the Act's provisions are binding and trigger heightened obligations for system providers.

---

[157] AI Act, art. 95(2), recitals 80, 165.

[158] Recital 27 of the AI Act provides that "[d]iversity, non-discrimination and fairness means that AI systems are developed and used in a way that includes diverse actors and promotes equal access." It also specifically references the 2019 Ethics Guidelines for Trustworthy AI. *See supra* note 68.

[159] AI Act, recitals 88–89.

[160] AI Act, recital 68.

[161] These third parties are not otherwise subject to the AI Act.

[162] AI Act, art. 2(12), recitals 89 and 102. Exceptions apply to open-source GPAI models with systemic risk.

[163] AI Act, Chapter VI, specifically Article 63.

[164] AI Act, recital 142.

[165] AI Act, art. 5(1)(c). This prohibition applies if the social scoring is based on "social behaviour or known, inferred or predicted personal or personality characteristics."

#### c. *High-Risk AI System Deployer Obligations*

Many of the equity-related implications of AI agents crystallize at the point of deployment, that is, the specific application that may affect relevant individuals and groups. It is here that the AI Act's primary equity-related instrument—the Fundamental Rights Impact Assessment (FRIA)—comes into play. Article 27 provides:

> 1. Prior to deploying [certain] high-risk AI system[s] referred to in Article 6(2), [...] deployers that are bodies governed by public law, or are private entities providing public services, and deployers of high-risk AI systems referred to in points 5 (b) and (c) of Annex III, shall perform [a FRIA] consisting of: […]
>
>> (c) the categories of natural persons and groups likely to be affected by its use in the specific context;
>> (d) the specific risks of harm likely to have an impact on [them], taking into account the information given by the provider pursuant to Article 13;
>> (f) the measures to be taken in the case of the materialisation of those risks, including the arrangements for internal governance and complaint mechanisms.

These provisions primarily cover high-risk AI systems used for creditworthiness or insurance risk assessment.[168] In these settings, deployers must notify Market Surveillance Authorities of the results and update the assessment when circumstances change.[169] In addition, where high-risk AI systems contribute to decisions with legal or other significant effects, deployers must provide affected persons with clear and meaningful explanations of the system's role in the decision-making process.[170]

[166] AI Act, art. 6(2), Annex III(3)–(8).

[167] AI Act, art. 6(2), Annex III(1)(a)–(c).

[168] Fundamental rights under the Charter of Fundamental Rights of the European Union include: the right to human dignity and life, the right to respect for private and family life, protection of personal data, freedom of thought, conscience, religion, expression, information, assembly and association, equality before the law and non-discrimination, the rights of the child and the elderly, access to justice and a fair trial, and social and economic rights such as the right to work and to health care.

[169] AI Act, art. 27(1)–(3).

The FRIA is an important mechanism for incorporating equity considerations into deployment decisions. Yet its scope is narrow. Many high-risk uses of AI with comparable equity implications, such as employment management systems, private educational AI systems, commercial biometric applications, and privately operated critical infrastructure, do not trigger a FRIA obligation under the Act. As a result, the mechanism's capacity to address unfair decision-making is limited to a narrow range of deployments.

Even where a FRIA is required, however, the mechanism may be a poor fit for addressing equity-related concerns from AI agents. A key issue is that a FRIA is a one-off or periodic exercise, while AI agents may regularly alter their behavior in ways that should warrant reassessment. Other deployer obligations do not address the issue. Human oversight requirements are framed at a relatively high level;[171] monitoring obligations depend largely on provider instructions;[172] and the requirement of automated logging does not explicitly focus on equity-related concerns.[173] Moreover, the requirement that deployers ensure input data are "sufficiently representative in view of the intended purpose"[174] seems to contemplate a static assessment at deployment, not the continuous and adaptive monitoring that is needed for AI agents.

Transparency obligations under the Act complement FRIAs by requiring that individuals be informed when they are exposed to certain AI systems, including high-risk decision-making systems, biometric categorization, emotion recognition and AI-generated content.[175] While these obligations do not impose substantive fairness or equity requirements, they play a critical role in informing affected persons and groups of AI systems with potentially detrimental outcomes.

---

[170] AI Act, art. 86.

[171] Article 26(2) of the AI Act only requires that deployers assign this task to competent personnel.

[172] A real-time auditing obligation may be found in Article 26(5), which requires the monitoring of the AI system's operation "on the basis of the instructions for use," but this obligation depends on the content of those instructions.

[173] AI Act, art. 26(6).

[174] Article 26(4) of the AI Act requires the deployer to ensure that "input data is sufficiently representative in view of the intended purpose," which is an ostensibly static requirement, although there is some room for a more dynamic interpretation.

[175] AI Act, arts. 26(11), 50(3)–(4).

#### d. *High-Risk AI System Provider Obligations*

While deployment decisions shape how equity-related risks materialize in practice, many of those risks are conditioned earlier, through design choices made by system providers. These choices influence how AI agents process information, whose data are reflected in their outputs, and how disparities may be amplified or constrained once systems are deployed.

Accordingly, several provider obligations under the AI Act may (indirectly) help address equity-related risks. Providers must, inter alia, take measures to support the development of AI literacy of their staff and other persons dealing with the operation and use of AI systems on their behalf,[176] supply deployers with comprehensible instructions enabling lawful and appropriate use,[177] and disclose information about system performance, including variations in performance across persons or relevant groups.[178] These obligations can shape whether equity-related impacts can be recognized, understood, and addressed in practice. In addition, high-risk AI systems must be registered in a central EU database, with additional registration obligations for public sector deployers. However, several important categories such as law enforcement and migration are considered non-public, which limits the external scrutiny of equity-relevant AI uses.[179]

Apart from these binding obligations, the AI Act recognizes broader equity concerns through Article 95, which invites the AI Office and Member States to facilitate the development of voluntary codes of conduct. These codes of conduct are intended to articulate ethical commitments applicable across AI systems and actors, including providers and deployers. The codes thus offer a flexible (albeit non-binding) avenue for addressing equity considerations that fall outside the Act's more prescriptive requirements.

The Act's most direct (and binding) engagement with equity-related concerns at the high-risk AI system provider level lies in its data governance requirements. Providers of high-risk AI systems must assess the availability, quantity, and suitability of datasets and examine whether biases are likely to affect health, safety, or fundamental rights, including prohibited discrimination.[180] To enable such assessment, the Act permits the (otherwise unlawful) processing of special categories of personal data for the purpose of bias detection and correction.[181] Special categories of data include information about a person's racial or ethnic background, political views, religious or philosophical beliefs, trade union membership, genetic or biometric identifiers used to uniquely identify a person, health information, and details about a person's sex life or sexual orientation.[182]

[176] Digital Omnibus on AI, art. 1(5), amending AI Act, art. 4.
[177] AI Act, art. 13(2)–(3) and recital 72.
[178] AI Act, art. 13(3)(b)(v).
[179] AI Act, arts. 49, 71, and recital 131.
[180] AI Act, art. 10(2)(f)–(g).

This mechanism, while appropriate for governing some conventional AI systems, does not necessarily map well onto AI agents. The mechanism assumes an order of operations in which sensitive data are briefly introduced for testing or correction, used to measure bias, and then removed.[183] For AI agents, however, bias detection and correction need to be an ongoing and iterative process. Even if this were required, on a practical level it would be very difficult to operationalize such requirements in AI agents that adapt over time and in response to new tasks and applications.[184]

#### e. *Model Provider Obligations*

Equity-related concerns also stem from the capacities and biases of GPAI models, which can then affect downstream systems and deployments. The obligations of model providers are particularly salient in this context. Notably, however, the obligations imposed on providers of GPAISR models apply only where equity-related concerns qualify as "systemic risks." This reveals a limitation in the AI Act's risk framework, which limits "systemic risk" to risks "specific to the high-impact capabilities" of the most advanced GPAI models, conflating what constitutes systemic risk with appropriate governance requirements.[185] As a result, equity-related concerns are not included among the specified systemic risks that necessarily trigger heightened obligations under the Act.

[181] Digital Omnibus on AI, art. 1(6), adding AI Act, art. 4a and replacing AI Act, art. 10(5), provides a legal basis for processing special categories of personal data beyond those in the GDPR, subject to enumerated conditions, when undertaken to identify and correct biases in high-risk AI systems. *See also* Marvin van Bekkum, *Using Sensitive Data to De-Bias AI Systems: Article 10(5) of the EU AI Act,* 56 COMP. L. & SEC. REV. 106115 (2025); European Parliamentary Research Service, *Algorithmic Discrimination under the AI Act and the GDPR* (Feb. 2025), https://www.europarl.europa.eu/RegData/etudes/ATAG/2025/769509/EPRS_ATA%282025%29769509_EN.pdf.

[182] GDPR, art. 9(1).

[183] Digital Omnibus on AI, art. 1(6), adding AI Act, art. 4a(1)(e) (previously AI Act art. 10(5)(e)).

[184] *See* van Bekkum, *supra* note 181.

[185] *See* Philipp Hacker et al., *AI, Digital Platforms, and the New Systemic Risk*, PROC. 2026 ACM CONF. ON FAIRNESS, ACCOUNTABILITY, AND TRANSPARENCY (2026).

This creates uncertainty as to whether risks of certain discriminatory outcomes will be treated as systemic risks under the Act, particularly where they arise from non-frontier models. Consider, for example, early studies of (non-frontier) language models that associate Muslim identity with violence[186] and LLM-based résumé screening systems that disadvantage candidates along gender and racial lines.[187] If such risks are considered to stem from downstream deployment rather than the models themselves, they may evade scrutiny at the stage where intervention would be most effective.

Where model level obligations fail to address equity-related concerns, the AI Act relies mainly on information and disclosure obligations of downstream system providers and deployers. While granular disclosure requirements do not explicitly address bias or discriminatory behavior of models,[188] a catch-all obligation requires GPAI model providers to supply downstream actors with the information necessary for their compliance.[189] Because deployers conducting FRIAs[190] depend on information provided by system providers,[191] who in turn rely on model providers, this obligation arguably requires that GPAI model providers assess and provide information regarding certain equity-related concerns.

Turning to the issue of equitably distributing the benefits of AI agents, the AI Act's regulation at the model level is effectively silent. Transparency obligations are limited to high-level disclosures of safety frameworks and systemic risk assessments.[192] They offer little to no external scrutiny of who can access the capabilities and benefits of AI agents or shed light on how agents risk concentrating power.

[186] Abubakar Abid et al., *Large Language Models Associate Muslims with Violence*, 3 NAT. MACH. INTELL. 461 (2021).

[187] Kyra Wilson & Aylin Caliskan, *Gender, Race, and Intersectional Bias in AI Resume Screening via Language Model Retrieval*, PROC. AAAI/ACM CONF. ON AI, ETHICS & SOC'Y (2024).

[188] AI Act, art. 53(1)(b)(ii) and Annex XII.

[189] AI Act, art. 53(1)(b)(i).

[190] AI Act, art. 27(1)(d).

[191] AI Act, art. 13(3)(b)(iii), (v).

[192] GPAI CoP, Commitment 10.

### E. *Oversight*

A further governance challenge arising from AI agents concerns human oversight. Meaningful human oversight requires ensuring that humans can monitor AI agents in real time and, where necessary, override their actions.[193] On a practical level, it is difficult to ensure effective oversight in circumstances where AI agents operate at superhuman speed and scale, including where multiple agents (and subagents) interact with one another.[194] Limited understanding of how AI agents reason and make decisions further hinders efforts to oversee them.[195]

Concretely, some traditional oversight tools, such as emergency kill-switches and rollbacks, might not be effective in the case of AI agents. Kill-switches, which involve shutting down a system upon certain triggering events, require that those events can be anticipated and specified in advance.[196] For AI agents that operate autonomously in novel scenarios, this will often not be the case. Rollbacks, meanwhile, require there to exist clearly defined "safe states" to which a system can be returned. For AI agents that take consequential and irreversible actions, this may be impossible.[197]

For completeness, it is also worth noting that even where effective oversight remains possible, establishing such oversight can impose significant costs on deploying AI agents and undermine their core value proposition, namely, the ability to operate autonomously.[198]

---

[193] *See generally* Sarah Sterz et al., *On the Quest for Effectiveness in Human Oversight: Interdisciplinary Perspectives*, PROC. 2024 ACM CONF. ON FAIRNESS, ACCOUNTABILITY, AND TRANSPARENCY 2495, 2495–2507 (2024); Alan Chan et al., *Visibility into AI Agents*, PROC. 2024 ACM CONF. ON FAIRNESS, ACCOUNTABILITY & TRANSPARENCY 958 (2024). *See also* European Data Protection Supervisor, *Human Oversight of Automated Decision-Making*, PUBLICATIONS OFFICE, EUROPEAN UNION (2025), https://www.edps.europa.eu/system/files/2025-09/25-09-15_techdispatch-human-oversight_en.pdf; Rebecca Crootof et al., *Humans in the Loop*, 76 VAND. L. REV. 429 (2023).

[194] *See* Hammond et al., *supra* note 77.

[195] *See* Kolt, *supra* note 4, at 42.

[196] *See generally* Nate Soares et al., *Corrigibility*, AAAI WORKSHOP ON AI & ETHICS (2015); Elliott Thornley, *The Shutdown Problem: An AI Engineering Puzzle for Decision Theorists*, 182 PHIL. STUD. 1653 (2024).

[197] *See* Alan Chan et al., *Infrastructure for AI Agents*, TRANSACTIONS ON MACHINE LEARNING RESEARCH (2025).

[198] *See* Kolt, *supra* note 4, at 35.

### 1. The AI Act's Response

Addressing these challenges, the AI Act establishes a framework that imposes obligations on high-risk AI system providers and deployers. The framework generally aims to preserve the possibility of human intervention throughout system use.

#### a. *High-Risk AI System Provider and Deployer Obligations*

Article 14 is the central provision that stipulates the need for human oversight. It establishes a functional, relatively non-prescriptive standard for high-risk AI system providers:[199]

> 1. High-risk AI systems shall be designed and developed in such a way, including with appropriate human-machine interface tools, that they can be effectively overseen by natural persons during the period in which they are in use.[200]

This provision requires system providers to either embed oversight measures directly into the system or specify measures for deployers to implement.[201] Yet, the Act imposes only minimal obligations on deployers to actually implement these measures. While deployers must follow providers' instructions for use, this general obligation does not clearly extend to establishing complex oversight mechanisms.[202] The monitoring requirement in Article 26, meanwhile, mandates only passive observation and suspension when risks to health, safety, or fundamental rights arise. It does not mandate implementation of the oversight measures providers are required to specify.[203] In other words, providers must design for oversight and instruct deployers on the measures to accomplish this, but deployers are not required to implement those measures in practice.

The only explicit obligation relating to oversight is the requirement that human oversight be carried out by competent natural persons who possess the technical knowledge, training and authority necessary to understand the functioning of the AI system and to take appropriate action when needed.[204] The content of this obligation is reinforced by the requirement to support the development of AI literacy of the staff operating the system. Taken together, this light-touch approach is question-begging given that deployers are the actors closest to the system's real-world activities.[205]

---

[199] AI Act, art. 16(a).
[200] AI Act, art. 14(1).
[201] AI Act, art. 14(3)(a)–(b).
[202] AI Act, art. 13(1).
[203] AI Act, art. 26(5).
[204] AI Act, art. 26(2)–(3).

For high-risk AI system providers, the Act is more prescriptive, specifying the design requirements in Article 14(4), which provides that:

> [T]he high-risk AI system shall be provided to the deployer in such a way that natural persons to whom human oversight is assigned are enabled, as appropriate and proportionate:
>
> (a) to properly understand the relevant capacities and limitations of the high-risk AI system and be able to duly monitor its operation, including in view of detecting and addressing anomalies, dysfunctions and unexpected performance;
>
> (c) to correctly interpret the high-risk AI system's output [...]
>
> (d) to decide [...] not to use the high-risk AI system or to otherwise disregard, override or reverse the output of the high-risk AI system;
>
> (e) to intervene in the operation of the high-risk AI system or interrupt the system through a 'stop' button or a similar procedure that allows the system to come to a halt in a safe state.

The primary problem here is that Article 14(4)'s oversight requirements presuppose a control framework that is difficult to establish for AI agents in practice. The listed obligations (naively) assume that relevant system behavior can be rendered legible to human overseers in real time and that halting and/or reversing the actions of AI agents is technically feasible.[206]

#### b. *Model Provider Obligations*

The AI Act and Code of Practice impose few specific requirements on GPAISR model providers concerning the oversight of AI agents. The Code recognizes loss of control as a specified systemic risk[207] and flags several oversight concerns as systemic risk sources. These include, for example, capabilities to evade human oversight[208] and the general level of oversight (or lack thereof).[209] Yet, the Code of Practice offers little concrete guidance on mitigation. The Code does not discuss specific interventions for controlling agents, such as oversight protocols or emergency stops at the model level.

[205] Kaminski & Selbst, *supra* note 8, at 1127.

[206] *See* Soares et al., *supra* note 196; Thornley, *supra* note 196.

[207] GPAI CoP, Appendix 1.4(2).

[208] GPAI CoP, Appendix 1.3.1(9).

[209] GPAI CoP, Appendix 1.3.3(3).

Beyond a brief reference to transparency into model reasoning, the framework remains largely silent on agent-specific oversight measures.[210]

At the same time, the oversight duties imposed on system providers under Article 14 are difficult to fulfill without meaningful insight into the underlying model. Obligations to help users interpret system outputs, detect failures, and understand system limitations all assume access to granular information regarding model behavior. The AI Act's disclosure requirements for model providers, however, are not sufficiently detailed to support such hands-on oversight of AI agents.[211] The result is that while system providers are expected to exercise robust oversight, they will likely lack access to the information needed to do so effectively.

* * * *

The preceding analysis has examined how the AI Act responds to five governance challenges posed by AI agents: ensuring reliable *performance*, preventing *misuse*, protecting *privacy*, promoting *equity*, and maintaining meaningful human *oversight*. Across these challenges, a consistent pattern emerges. The Act addresses many governance challenges through its risk classification mechanisms, substantive requirements for providers and deployers, and information obligations along the AI value chain. Yet, these mechanisms are poorly suited to agents whose behavior changes over time as they operate in new environments and interact with other systems. The effectiveness of the AI Act's response, however, turns not only on the scope and content of its substantive requirements but also on the institutions that administer and enforce those requirements.

## III. INSTITUTIONAL IMPLEMENTATION

The institutional implementation of the AI Act and associated policy instruments relies on a combination of industry-led *self-regulation* and public *enforcement* by designated authorities, both of which are shaped by the availability of governance *resources*. This Part examines each of these elements in turn.

---

[210] GPAI CoP, Measure 5.1(8).

[211] AI Act, art. 53(1)(b)(i).

## A. *Self-Regulation*

Industry self-regulation plays a central role in the EU AI Act, both in its design and in its operation. The Act relies heavily on technical standards and Codes of Practice to translate its high-level requirements into concrete obligations, and it assigns industry actors a substantial role in shaping both. While industry participation is a familiar feature of EU product regulation, the AI Act places unusually strong weight on these self-regulatory instruments compared to earlier regulatory frameworks.[212] Even with such mechanisms in place, AI providers retain significant discretion in how obligations are interpreted and implemented. This discretion is especially consequential for AI agents, as existing standards do not (yet) adequately reflect agent-specific characteristics. Similar gaps arise in risk classification, where the Act offers limited guidance on how to account for the multi-purpose and adaptive nature of AI agents.

Technical standards under the AI Act are developed through CEN/CENELEC JTC-21, the relevant European standardization body,[213] whose membership is largely drawn from industry.[214] Although participation is formally open to actors from academia and civil society, in practice such involvement is limited, often due to resource constraints or lack of awareness.[215] The European Commission may intervene only if the standardization process fails, by issuing "common specifications"[216] that provide binding technical guidance. At the time of writing, the standards intended to clarify obligations for high-risk AI systems are expected by the fourth quarter of 2026.[217]

---

[212] *See* European Commission, *New Legislative Framework* (2008), https://single-market-economy.ec.europa.eu/single-market/goods/new-legislative-framework_en.

[213] AI Act, art. 40.

[214] *See* Robert Kilian et al., *European AI Standards – Technical Standardisation and Implementation Challenges under the EU AI Act*, 16 EUR. J. RISK REG. 1038, 1038–1062 (2025).

[215] *Id.*

[216] AI Act, art. 41.

[217] *See* European Commission, *Commission Implementing Decision of 22 May 2023 on a Standardisation Request to the European Committee for Standardisation and the European Committee for Electrotechnical Standardisation in Support of Union Policy on Artificial Intelligence* C(2023)3215 final, art. 3 (May 22, 2023), https://ec.europa.eu/transparency/documents-register/detail?ref=C(2023)3215&lang=en. This delay in the standardization process was a key factor in delaying the entry into force of the high-risk AI system obligations under the AI Act. *See* Digital Omnibus on AI, recital 2.

The GPAI Code of Practice is also shaped to a significant extent by industry participation. The concept of general-purpose AI was introduced only late in the AI Act's legislative process,[218] leaving the associated provisions relatively underspecified. Because formal standardization processes were not well suited to operationalize these obligations within the AI Act's short implementation timeline, the legislature opted to supplement them with a Code of Practice.

Published in July 2025, the GPAI Code of Practice followed a nine-month process involving more than a thousand stakeholders from industry, academia, and civil society.[219] Although the process was chaired by independent experts and included a broad range of contributors, the AI Act assigns a leading role to GPAI model providers, with other stakeholders occupying a more limited, supportive role.[220] As the drafting process neared completion, concerns were raised that several leading U.S.-based model providers enjoyed privileged access that resulted in disproportionate influence over the final text.[221] At the time of writing, 24 model providers have signed the Code of Practice, including OpenAI, Anthropic, and Google, while Meta remains a prominent holdout, continuing to publicly oppose the instrument.[222]

Both the standards for high-risk AI systems and the Code of Practice grant providers considerable discretion in how compliance obligations are implemented. Consider, for example, a provider who consults the relevant technical standard regarding the expected accuracy of an AI agent under the Act. While the standard sets out general accuracy metrics and methodological guidance, it does not specify concrete benchmarks or thresholds at which a system can be considered sufficiently accurate for market placement.[223]

---

[218] *See* Schwartmann & Zenner, *supra* note 20, at 3–9.

[219] *See* European Commission, *General-Purpose AI Code of Practice Now Available* (Jul. 10, 2025), https://digital-strategy.ec.europa.eu/en/news/general-purpose-ai-code-practice-now-available.

[220] AI Act, art. 56(3).

[221] *See* Paul Nemitz & Amin Oueslati, *How US Firms Are Weakening the EU AI Code of Practice*, TECHPOLICY.PRESS (June 30, 2025), https://www.techpolicy.press/how-us-firms-are-weakening-the-eu-ai-code-of-practice/.

[222] *See* Euronews, *Meta Rebuffs EU's AI Code of Practice* (Jul. 18, 2025), https://www.euronews.com/next/2025/07/18/meta-rebuffs-eus-ai-code-of-practice.

[223] *See* Josep Soler Garrido et al., *Harmonised Standards for the European AI Act*, SCIENCE FOR POLICY: INFORMATION SOCIETY STANDARDS, EUROPEAN COMMISSION (Oct. 24, 2024), https://publications.jrc.ec.europa.eu/repository/handle/JRC139430.

A similar issue arises in the Code of Practice. Rather than prescribing particular evaluation methods or mitigation techniques, it requires providers to select measures they consider "appropriate" or consistent with the "state of the art," and to justify those choices.[224] The Code of Practice gives effect to this approach by listing broad categories of possible safeguards, such as mechanisms to support safer agent ecosystems, including system identification methods, communication protocols, or incident monitoring tools. The Code, however, does not actually mandate their adoption.[225] In practice, the implementation of such measures will depend on the provider's own assessment of what is "suitable and necessary."[226] While this flexibility may enhance governance adaptability over time, it also invites divergent interpretations and complicates enforcement.

Relatedly, the AI Act does not require regulatory approval of AI systems prior to market placement. Instead, it relies primarily on providers to assess conformity in advance,[227] including classifying their systems' risk level; identifying applicable obligations; and determining whether those obligations are met.[228] External conformity assessments are required only in limited circumstances.

This framework places substantial interpretive responsibility on regulated actors. Determining whether a system's intended purpose falls within a high-risk category is often far from straightforward. Empirical evidence underscores this difficulty. In an expert-led study examining 106 different AI use cases, nearly 40 percent could not be conclusively classified under existing risk categories.[229] Similar ambiguities arise for AI agents. For example, it is unclear whether an AI agent used for on-the-job training within a company should be treated as part of an "educational or vocational training institution" and thus classified as high-risk.[230] Likewise, a legal AI agent that analyzes leasing contracts may, in the context of litigation, perform functions akin to evaluating evidence for a judicial authority, potentially bringing it within the scope of high-risk regulation.[231]

[224] GPAI CoP, Commitments 5–7.

[225] GPAI CoP, Measure 5.1.

[226] GPAI CoP, Glossary.

[227] AI Act, art. 43, recital 125.

[228] AI Act, art. 6(4).

[229] *See* AppliedAI Initiative, *AI Act: Risk Classification of AI Systems from a Practical Perspective* (Mar. 2023), https://www.appliedai.de/en/insights/ai-act-risk-classification-of-ai-systems-from-a-practical-perspective/.

[230] AI Act, Annex III(3)(a).

## B. *Enforcement*

The AI Act's enforcement regime for AI agents involves a complex allocation of authority between (a) national Market Surveillance Authorities ("MSAs"), which are existing national bodies responsible for supervising product compliance and market safety; and (b) the European Commission's AI Office, a newly created EU-level body tasked with overseeing GPAI models and systemic risks.[232]

Enforcement of high-risk AI system obligations generally falls to national MSAs under pre-existing market surveillance rules, with additional AI Act provisions applying in parallel. Enforcement of obligations pertaining to GPAI models, meanwhile, falls to the AI Office.[233] Where an AI system is based on a GPAI model developed by the same entity, enforcement responsibility shifts to the AI Office, which is granted some of the powers held by MSAs under the Market Surveillance Regulation ("MSR").[234] However, EU-wide enforcement measures still require coordination involving national authorities, including, for example, where taking coercive measures on national territory.[235]

Authorities identify non-compliance or serious risks through three primary channels: (1) mandatory information obligations; (2) external alerts; and (3) active investigations. Information obligations of GPAI model providers under the AI Act require staged disclosure during the model's development and deployment. First, providers of GPAI models with systemic risk must notify the AI Office when they anticipate that a model under training will cross the threshold for systemic risk, for example, when scaling the computational resources used in training.[236] Second, providers must submit a Safety and Security Framework under the Code of Practice, setting out how they intend to identify, assess, and mitigate risks such as loss of control or harmful manipulation in foreseeable uses.[237] Finally, by the time a model is placed on the market, providers must supply a model-specific "Safety and Security Model Report," documenting the results of evaluations and the mitigation measures actually adopted for that particular model version.[238]

---

[231] AI Act, Annex III(8)(a).

[232] AI Act, recital 161.

[233] AI Act, arts. 70, 74, 88.

[234] AI Act, art. 75a, introduced by Digital Omnibus on AI, art. 1(31).

[235] *See* Michiel Luchtman, *Setting the Scene: The Rise of EU Law Enforcement Authorities*, *in* EU ENFORCEMENT AUTHORITIES: PUNITIVE LAW ENFORCEMENT IN A COMPOSITE LEGAL ORDER 7, 13 (Michiel Luchtman et al. eds., 2023) (suggesting that the Commission "cannot exercise physical coercion itself in case of non-cooperation. For that, it needs the assistance of national authorities. It does have the power, however, to impose administrative sanctions, including punitive sanctions, for non-cooperation. Coercion, therefore, is not a Commission power, but compulsion is.").

For high-risk AI systems, providers must register in the EU database before market placement, including declarations of conformity,[239] while deployers must register summaries of Fundamental Rights Impact Assessments and Data Protection Impact Assessments (both discussed above).[240] These disclosure requirements are intended to equip regulators with early visibility into developments that may warrant supervisory attention.

Both high-risk AI system providers and GPAI model providers are required to report serious incidents where a causal link to the AI system or model can be established.[241] For AI agents, establishing that link can be difficult, because harmful outcomes may result from the interaction of multiple components and systems—such as the underlying model, external tools, and deployment settings—rather than from a single identifiable point of failure.[242] As a result, determining what went wrong, who bears responsibility, and which data are needed for a meaningful root-cause analysis can be challenging. This is particularly the case where relevant information is distributed across different actors, including where the incident involves multiple systems developed or deployed by different entities.[243]

---

[236] AI Act, art. 52(1).

[237] GPAI CoP, Measure 1.4. The Framework (and updates thereto) must be provided within 5 business days of it being confirmed.

[238] GPAI CoP, Measure 7.7. The Model report notification can be delayed by up to 15 business days provided the Signatory is acting in good faith.

[239] AI Act, arts. 49(1), 71(2) and Annex VIII Section A.

[240] AI Act, art. 49(3), Annex VIII Section C.

[241] AI Act, arts. 73(1)–(2), 55(1)(c), GPAI CoP, Commitment 9. At the time of writing, draft guidance has been provided, setting out the requirements under Article 73 of the AI Act in greater detail. *See* European Commission, *AI Act: Commission Issues Draft Guidance and Reporting Template on Serious AI Incidents, and Seeks Stakeholders' Feedback* (Sept. 26, 2025), https://digital-strategy.ec.europa.eu/en/consultations/ai-act-commission-issues-draft-guidance-and-reporting-template-serious-ai-incidents-and-seeks.

[242] Carson Ezell et al., *Incident Analysis for AI Agents*, PROC. 8TH AAAI/ACM CONF. ON AI, ETHICS & SOC'Y (2025).

[243] European Data Protection Supervisor, *supra* note 193.

Following an incident, AI system and GPAISR model providers are required to conduct an incident investigation.[244] While this process can help support organizational learning and risk mitigation,[245] there is no specific requirement that AI system providers report the result of the incident investigation. The Code of Practice, however, requires more structured reporting processes from GPAISR model providers.[246] In terms of a regulatory response, only in the case of serious incidents involving high-risk AI *systems* does the AI Act explicitly task the competent authority with taking responsive measures, such as market recall.[247] By contrast, comparable intervention in relation to GPAISR *models* is left to the Commission's discretion,[248] with GPAISR model providers being required to propose regulatory actions in response to a serious incident.[249]

External parties complement these information obligations through four mechanisms: (1) qualified alerts from the Scientific Panel identifying Union-level risks;[250] (2) whistleblower reports protected under Directive 2019/1937;[251] (3) citizen complaints to MSAs;[252] and (4) downstream provider complaints to the AI Office concerning suspected infringements by GPAI model providers.[253]

Beyond passively receiving information, authorities possess extensive investigative powers to actively identify non-compliance and risks from AI systems. MSAs supervise testing in AI regulatory sandboxes and oversee real-world testing.[254] Authorities may evaluate AI systems presenting risks,[255]

---

[244] GPAI CoP, Measure 9.2(5), (8); AI Act, art. 73(6).

[245] AI Act, art. 73(6); GPAI CoP, Measures 9.2, para. 2 and 1.2, para. 2(2). *See generally* Isabel Richards et al., *From Incidents to Insights: Patterns of Responsibility Following AI Harms*, PROC. 5TH ACM CONF. ON EQUITY AND ACCESS IN ALGORITHMS, MECHANISMS, AND OPTIMIZATION 151 (2025).

[246] GPAI CoP, Measure 9.2(8).

[247] AI Act, art. 73(8).

[248] AI Act, art. 93(1)(c).

[249] GPAI CoP, Measure 9.2(7).

[250] AI Act, art. 90.

[251] Recital 172 of the AI Act explicitly provides that persons acting as whistleblowers in relation to infringements of the Regulation are protected under Union law.

[252] AI Act, art. 85, recital 170.

[253] AI Act, art. 27(1)(f).

[254] AI Act, art. 76(2).

[255] AI Act, art. 79(2).

evaluate systems classified as non-high-risk by providers,[256] perform checks on compliance,[257] demand full access to documentation and datasets,[258] and under certain conditions access source code.[259] The MSR further empowers authorities to require documentation from economic operators,[260] conduct unannounced on-site inspections and physical checks,[261] enter business premises,[262] initiate investigations,[263] and acquire product samples for inspection and reverse-engineering.[264] Similar powers are granted to the AI Office to monitor and investigate compliance of GPAISR model providers on whose models AI agents are built.[265]

Once an AI Act infringement has been identified, enforcement measures available to MSAs or the AI Office, depending on the applicable provision, include, among other things: requesting/requiring the provider to withdraw, recall, or restrict access to the market of the system or model,[266] taking corrective actions to cure a non-compliance,[267] affixing warnings,[268] and imposing fines.[269] Under certain circumstances, authorities are tasked with "ensuring" that a system is withdrawn, recalled, or its market access is restricted,[270] which can include requesting information society service providers or third parties to restrict access to APIs through which AI systems are accessed.[271]

Notably absent from the AI Act is a private right of action for EU citizens. Lawsuits, however, can be brought under other applicable EU law when individuals' rights are violated. Yet, combined with the limited public information on high-risk AI systems[272] and GPAI models[273], the overall result is that the responsibility for supervision and enforcement falls almost entirely on the EU AI Office and MSAs.

[256] AI Act, art. 80(1).
[257] AI Act, art. 80(8).
[258] AI Act, art. 74(12).
[259] AI Act, art. 74(13).
[260] MSR, art. 14(4)(a)–(c).
[261] MSR, art. 14(4)(d).
[262] MSR, art. 14(4)(e).
[263] MSR, art. 14(4)(f).
[264] MSR, art. 14(4)(j).
[265] AI Act, arts. 89, 91 and 92.
[266] AI Act, art. 93(1)(c); MSR, arts. 14(4)(h), 16(5).
[267] MSR, arts. 14(4)(g), 16(2)–(3); AI Act, art. 93(1)(a).
[268] MSR, arts. 14(4)(k)(i), 16(3)(g).
[269] MSR, arts. 14(4)(i), 41; AI Act, arts. 99, 101.
[270] MSR, art. 16(5).
[271] MSR, art. 14(4)(k)(ii).
[272] AI Act, arts. 49 and 71.
[273] GPAI CoP, Transparency Chapter, Measure 1.2.

### C. *Resourcing*

The efficacy of the AI Act's enforcement framework ultimately depends on the institutional capacity of the authorities tasked with implementing it. Insufficient government resources or expertise asymmetries vis-à-vis the private sector challenge both the development of sound regulation and its day-to-day enforcement. As discussed, effective rule-making and enforcement with respect to AI agents require sufficient talent not only in the AI Office but also within the relevant authorities at the EU Member State level. While governments generally struggle to recruit technical specialists, few domains face fiercer competition for top talent than AI.[274]

This human resources gap is particularly acute with respect to expertise in AI agents. Due to their novelty, the talent required for effectively evaluating AI agents is even scarcer. Despite the AI Office's initial recruiting successes, it is likely understaffed and fails to offer a competitive salary package to prospective talent.[275] These issues are reinforced by several structural barriers, such as the AI Office being subject to the European Commission's rigid hiring and employment rules. For example, reporting on the European Commission's AI Office highlights how slow recruitment processes, internal administrative requirements, and pressures to balance representation across Member States have hindered the office's ability to fill open positions, including leadership roles.[276] EU Member States face similar obstacles. Moreover, the simultaneous efforts of Brussels and Member States to build AI oversight functions further intensify competition for scarce resources.

As of late 2025, the EU AI Office reportedly employs 125 staff,[277] spanning technology specialists, operations personnel, lawyers, policy analysts, and economists, falling just short of its 140 full-time equivalent (FTE) goal.[278]

---

[274] *See Experts Call for Regular Reviews, Risk Priorities for EU's Code for AI Models* (Jul. 14, 2025), https://www.mlex.com/mlex/articles/2364139/experts-call-for-regular-reviews-risk-priorities-for-eu-s-code-for-ai-models.

[275] *See* Peder Schaefer, *Hiring Struggles Are Plaguing the EU AI Office*, TRANSFORMER (Sept. 15, 2025), https://www.transformernews.ai/p/eu-is-struggling-to-hire-ai--act-office-safety-unit.

[276] *Id.*

[277] *See* European Commission, *European AI Office* (Nov. 19, 2025), https://digital-strategy.ec.europa.eu/en/policies/ai-office.

[278] *See Enforcement of the EU AI Act: The EU AI Office* (June 12, 2024), https://cms-

These staff work across six units, including "Excellence AI and Robotics" and "AI for Societal Good." The enforcement of AI agent rules falls within the remit of the "AI Regulation and Compliance" and "AI Safety" units, although the exact division of responsibility remains unclear.

Concerns regarding whether the AI Office possesses sufficient expertise to effectively govern AI agents are brought into sharp relief when it is compared to specialized private organizations. For instance, the risk evaluations of METR, a leading AI evaluation organization, focus almost exclusively on agentic risks from frontier models, i.e., those that would likely be considered GPAISR models under the AI Act.[279] Despite this relatively narrow focus, METR alone employs 30 staff members.[280] This headcount highlights a stark resource disparity. If a single organization requires such significant talent to evaluate just one subset of agentic behavior, the AI Office's ability to provide comprehensive oversight across the full spectrum of agent risks appears increasingly tenuous.

The independent experts who led the drafting of the Code of Practice questioned the EU's ability to effectively enforce its rules given insufficient resources. They proposed increasing the number of staff in the AI safety unit, which is responsible for enforcing GPAI rules, to 100 FTE and expanding the full AI Act implementation team to 200 FTE. At the time of writing, Member States are still finalizing the structure of their MSAs, including their staff size. In Germany's most recent draft law, it proposed 100 FTE for its national body.[281] Beyond enforcing the AI Act's rules, their expanded portfolio will encompass a range of initiatives, including the operation of a regulatory sandbox, which allows AI systems to be tested under regulatory supervision before or during deployment, and the management of a dedicated service desk to support regulated entities with compliance.

---

lawnow.com/en/ealerts/2024/06/enforcement-of-the-eu-ai-act-the-eu-ai-office.

[279] *See* METR, *About METR*, https://metr.org/about.

[280] *Id.*

[281] Detailed staffing projections are contained in the full version of the following document: Bundesministerium für Digitales und Verkehr, *Referentenentwurf des Bundesministeriums für Digitales und Verkehr: Entwurf eines Gesetzes zur Durchführung der Verordnung (EU) 2024/1689 des Europäischen Parlaments und des Rates vom 13. Juni 2024 zur Festlegung harmonisierter Vorschriften für künstliche Intelligenz (KI-Verordnungs-Durchführungsgesetz — KIVO-DG)* (Sep. 11, 2025), https://bmds.bund.de/fileadmin/BMDS/Dokumente/Gesetzesvorhaben/CDR_Anlage1-250911_RefE_KIVO-Durchf%C3%BChrungsgesetz_Entwurf_barrierefrei.pdf.

Competitive compensation is a major barrier to recruiting these staff. Concretely, consider the gap in base compensation for senior technical roles in the EU AI Office ($52k to $109k) and roles in the UK AI Security Institute ($126k to $175k). This gap is even more pronounced when compared to private evaluation organizations. For instance, METR offers salaries for technical roles, such as research engineers, ranging from $200k to $340k. These figures, to be sure, are dwarfed by salaries at leading AI companies, where senior technical experts in European offices can command salaries between $241k and $422k, with some roles exceeding $724k[282]—without even accounting for equity-based compensation.[283] While the AI Office clearly cannot match private sector compensation, its current salary bands fail to achieve basic parity with peer public institutions.

The AI Office's strategy for obtaining computational resources ("compute") remains unclear, despite their importance in conducting evaluations of frontier AI systems. In addition, such evaluations can present security risks as they may involve stress-testing powerful models and using sensitive data, for which reliance on commercial cloud providers could be problematic.[284] At present, the AI Act provides no clear guidance on how the AI Office is expected to obtain access to the necessary (and significant)[285] computing

[282] Salary information for the EU AI Office (Brussels) is based on official European Commission salary grades. *See* European Commission, *2024 Annual Update of the Remuneration and Pensions of the Officials and Other Servants of the European Union and the Correction Coefficients Applied Thereto*, C/2025/2153, 2025 O.J. (C 2153) 11 (Apr. 11, 2025), https://eur-lex.europa.eu/legal-content/EN/TXT/?uri=CELEX%3A52025XC02153. Salary information for UK AISI (London) was obtained from published civil service pay bands and role-specific salary ranges available on the official AISI careers website and other UK government sources. Average salary bands for leading AI companies (OpenAI, Anthropic, Meta, Google) were constructed from salary reports on Glassdoor and Levels.fyi, filtering for European offices (primarily London, Dublin, and Munich). All figures were converted to U.S. dollars.

[283] Comparable discrepancies exist for policy and legal roles. For policy and legal officers, base compensation at the EU AI Office remains within the $52k to $109k range. By comparison, while UK AISI offers slightly lower starting salaries for policy roles ($49k to $74k) and slightly higher salaries for legal roles ($58k to $101k), both are significantly lower than in leading AI companies. In those companies' European offices, policy and legal roles command base salaries ranging from approximately $199k to $290k.

[284] To the extent the AI Office receives "white box," "gray box," or other privileged forms of access, security vulnerabilities may enable novel threat vectors, as well as raise concerns regarding intellectual property protection. *See* Alejandro Tlaie & Jimmy Farrell, *Securing External Deeper-Than-Black-Box GPAI Evaluations*, ARXIV (Mar. 13, 2025), https://arxiv.org/abs/2503.07496.

[285] *See*, *e.g.*, UK AI Security Institute, *Evidence for Inference Scaling in AI Cyber Tasks: Increased Evaluation Budgets Reveal Higher Success Rates* (Mar. 5, 2026),

resources and infrastructure. By contrast, the UK AI Security Institute has dedicated and priority access to national supercomputing resources.[286]

## IV. Lessons Learned

Our systematic analysis of the EU AI Act's regulatory and institutional response to the governance challenges arising from AI agents offers important lessons for lawmakers and technologists both within the European Union and in other jurisdictions. We focus on three themes: *artifact-centric regulation* that seeks to govern technical components of AI systems rather than shape how agents operate in real-world settings; the *many-hands problem* that arises when responsibility for managing risk is spread across multiple actors; and *institutional capacity* constraints that undermine the ability to effectively monitor the actions of AI agents and prevent resulting harm.

### A. *Artifact-Centric Governance*

The EU AI Act regulates AI primarily by reference to discrete technical artifacts—models and systems—whose risks are assumed to be clearly identifiable, attributable to particular actors, and amenable to mitigation.[287] AI agents challenge this approach. The actions of AI agents, and thus the risks they pose, depend on the environments in which they operate, the tools they can access, and the permissions they are granted.[288]

These difficulties are most salient in the AI Act's approach to risk classification, which determines when the Act's most demanding obligations apply. For high-risk AI systems, the AI Act relies heavily on *ex ante* classifications tied to a system's intended use.[289] For model level governance, the AI Act relies on systemic risk designations linked primarily to advanced capabilities approximated by compute thresholds.[290] As the preceding analysis has shown, this framework is not appropriate for agentic systems

https://www.aisi.gov.uk/blog/evidence-for-inference-scaling-in-ai-cyber-tasks-increased-evaluation-budgets-reveal-higher-success-rates.

[286] *See* Department for Science, Innovation and Technology, *UK Compute Roadmap* (Jul. 17, 2025), https://www.gov.uk/government/publications/uk-compute-roadmap/uk-compute-roadmap.

[287] *See* Kaminski & Selbst, *supra* note 8; Almada & Petit, *supra* note 8; Leufer & Hidvégi, *supra* note 8; Gstrein, *supra* note 16; Hacker et al., *supra* note 16.

[288] *See* Carl Gahnberg, *What Rules? Framing the Governance of Artificial Agency*, 40 POLICY & SOC. 194, 196, 203–4 (2021).

[289] *See*, *e.g., supra* Parts I.A and II.C.

[290] AI Act, art. 51(2). *See also* Matteo Pistillo et al., *The Role of Compute Thresholds for AI Governance*, 1 GEO. WASH. J. L. & TECH. 29 (2025).

whose risks are neither static nor fully determined at the time of model training or market placement. For AI agents, risks often emerge in new deployment contexts or as agents are granted new affordances, which may have little connection to their predefined intended purpose. In such cases, heightened risks are not met with heightened regulatory obligations.

There are, however, alternative approaches. Recent U.S. state legislation, namely, California's Transparency in Frontier Artificial Intelligence Act[291] and New York's Responsible AI Safety and Education (RAISE) Act,[292] shift the trigger for heightened obligations away from the *technical artifact* and toward characteristics of the *developer*. This approach avoids some of the foregoing risk classification challenges under the EU AI Act. However, the two aforementioned legislative instruments do so by narrowing their scope of regulation to large developers—an approach that may fail to contend with risks arising from highly capable AI agents developed or deployed by smaller companies or organizations.

To be sure, the challenge of regulating technologies whose risks cannot be meaningfully characterized by reference to a discrete technical artifact alone is not unique to AI agents. Foundational work on technology law and governance cautions against treating complex, distributed systems as static and isolated products.[293] Regulation aimed exclusively at technical artifacts tends to miss critical institutional, organizational, and environmental conditions. The regulation of AI agents is no exception.

These observations invite a two-step shift in perspective. First, mechanisms for governing AI agents should look outward toward the sociotechnical environments in which agents operate, recognizing that agents' behavior is shaped by the tools, permissions, and interfaces to which they have access.[294] Second, governance mechanisms should turn to areas of extant law that already regulate and structure those environments and resources, such as contract law, tort law, and financial regulation.[295] We hope to study these in future work.

---

[291] Transparency in Frontier Artificial Intelligence Act, Cal. Bus. & Prof. Code §§ 22757.10-16; Cal. Gov't Code § 11546.8; Cal. Lab. Code § 1107.

[292] Responsible AI Safety and Education Act, N.Y. Gen. Bus. Law §§ 1420–29.

[293] *See* DAVID COLLINGRIDGE, THE SOCIAL CONTROL OF TECHNOLOGY 13 (1980); Rebecca Crootof & BJ Ard, *Structuring Techlaw*, 34 HARV. J.L. & TECH. 347, 357–59 (2021).

[294] *See* Ludovic Terren, *Beyond Safe Models: Why AI Governance Must Tackle Unsafe Ecosystems*, TECHPOLICY.PRESS (May 1, 2025), https://www.techpolicy.press/beyond-safe-models-why-ai-governance-must-tackle-unsafe-ecosystems/; Chan et al., *supra* note 193; Andrew W. Torrance & Bill Tomlinson, *Agents in a Tangled Bank: An Ecosystem Approach to AI Regulation*, 20 FIU L. REV. 567 (2025).

### B. *The Many-Hands Problem*

Our analysis of the EU AI Act illustrates that many of the governance challenges stemming from AI agents are exacerbated by their distributed development, deployment, and operation. Rather than being attributable to a single entity, the actions of agentic systems are often shaped by multiple actors and resources. The result is a "many-hands problem" in which responsibility, control, and knowledge are fragmented, leaving no participant with a complete view of, or responsibility for, the resulting risks.[296]

The AI Act's value chain approach recognizes this fragmentation in principle by allocating obligations across different roles. In practice, however, the Act assumes that downstream actors can identify and manage risks on the basis of upstream assurances and disclosures. This assumption may fail where system providers or deployers are tasked with assessing agent-specific risks without access to sufficiently timely or granular information. Effective governance in such settings depends on access to technical resources that would allow downstream actors to monitor or override agents' actions. Without such resources, actors subject to formal governance responsibilities would lack the practical tools needed to comply with their obligations.

Multi-agent settings in which different AI agents communicate and interact with one another compound this challenge.[297] An effective governance response requires, at the very least, ecosystem-level information that no single actor can obtain on its own.[298] By implementing a narrowly scoped role-based approach, the AI Act clearly falls short. Pooling key information and insights already collected pursuant to the AI Act into the EU AI Office or another centralized body would be a useful starting point.

---

[295] *See* CHOPRA & WHITE, *supra* note 7; CHINEN, *supra* note 7; TURNER, *supra* note 7; ABBOTT, *supra* note 7; CHESTERMAN, *supra* note 7.

[296] *See* Helen Nissenbaum, *Accountability in a Computerized Society*, 2 SCI. & ENG'G ETHICS 25 (1996); Mark Coeckelbergh, *Artificial Intelligence, Responsibility Attribution, and a Relational Justification of Explainability*, 26 SCI. & ENG'G ETHICS 2051 (2020); A. Feder Cooper et al., *Accountability in an Algorithmic Society: Relationality, Responsibility, and Robustness in Machine Learning*, PROC. 2022 ACM CONF. ON FAIRNESS, ACCOUNTABILITY & TRANSPARENCY 864 (2022).

[297] *See* Hammond et al., *supra* note 77.

[298] *See* THE AI AGENT INDEX, *supra* note 5; Staufer et al., *supra* note 5.

Perhaps the thorniest "many hands problem" affecting AI agents concerns entities that altogether fall outside the AI Act's core regulatory categories. Specifically, the tools used by AI agents that are central to their capabilities and risks are often developed or maintained by third-party providers, which are considered neither GPAI model providers nor AI system providers under the AI Act. Consequently, their governance is highly limited. In such cases, the Act relies primarily on contractual arrangements to support information exchanges and risk management processes between high-risk system providers and component suppliers. Although the AI Office may develop voluntary model contractual terms supporting such arrangements,[299] it remains unclear whether this mechanism can facilitate the information flows and technical cooperation needed to effectively govern AI agents.

### C. *Institutional Monitoring*

The final area we consider concerns the institutional capacity of regulators to monitor AI agents after their deployment. As discussed, many of the risks associated with agents only become apparent in the specific contexts in which they are deployed, and then continue to change over time as agents learn and adapt to new tasks and environments. The EU AI Act, however, largely relies on reporting that takes place prior to market placement. As a result, while EU authorities may receive information that is accurate at the time of reporting, that information may fail to capture the most consequential aspects of agents' real-world impact.

Such limited monitoring also weakens provisions of the AI Act that depend on up-to-date technical judgments. Most notably, the requirement that model evaluations be assessed against the "state of the art"[300] will be difficult to enforce if regulators do not have timely access to information about current agent capabilities and risks. Moreover, while the AI Act requires providers to update risk assessments and technical documentation, these obligations can generally be fulfilled through periodic reports, the contents of which might (once again) fail to enable regulators to effectively monitor agents.

A more robust approach to monitoring AI agents might focus less on one-off reporting requirements and more on maintaining ongoing visibility into how these systems behave in practice,[301] including leveraging ongoing research into technical methods for monitoring AI agents.[302] Stepping back, the monitoring of AI agents is clearly not a legislative drafting problem that can be addressed solely through more detailed or different rules. As with other challenges arising from AI agents, effective regulatory responses will need to be accompanied by appropriate institutional capacity, resourcing, and technical infrastructure.

---

[299] AI Act, art. 24(4), para. 2.

[300] GPAI CoP, Measure 3.2.

[301] Chan et al., *supra* note 193.

## CONCLUSION

The governance challenges presented by AI agents do not primarily concern regulatory scope, but regulatory fit. Existing regulatory frameworks such as the EU AI Act apply to AI agents in principle, but fall short in practice. In particular, the AI Act's assumption that societal risks can be traced to a single technical artifact, assessed at a fixed point in time, and attributed to a predefined set of actors is misguided. Risks from AI agents arise through interactions with a growing array of complex tools, actors, and environments. To contend with these problems, lawmakers in the EU and other jurisdictions will need to look beyond refining current legislative instruments and toward expanding the technical expertise and operational capacity of regulators. Seen in this light, AI agents not only challenge existing efforts to regulate AI, but invite us to reimagine how regulation should contend with a new and rapidly evolving class of autonomous systems.

[302] *See*, *e.g.*, Bowen Baker et al., *Monitoring Reasoning Models for Misbehavior and the Risks of Promoting Obfuscation*, ARXIV (Mar. 14, 2025), https://arxiv.org/abs/2503.11926; Korok Ray, *Monitoring Teams of AI Agents*, 84 J. ARTIF. INTELL. RES. 1 (2025); Anita Rao et al., *Challenges to the Monitoring of Deployed AI Systems*, CENTER FOR AI STANDARDS AND INNOVATION, NATIONAL INSTITUTE OF STANDARDS AND TECHNOLOGY (Mar. 2026), https://nvlpubs.nist.gov/nistpubs/ai/NIST.AI.800-4.pdf.